\documentclass[fleqn,usenatbib]{rasti}

\usepackage{newtxtext,newtxmath}

\usepackage[T1]{fontenc}

\DeclareRobustCommand{\VAN}[3]{#2}
\let\VANthebibliography\thebibliography
\def\thebibliography{\DeclareRobustCommand{\VAN}[3]{##3}\VANthebibliography}

\usepackage{graphicx}	
\usepackage{amsmath}	

\title[Radio emission of TDEs]{Radio emission of tidal disruption events from wind-cloud interaction}

\author[De-Fu Bu et al.]{
De-Fu Bu,$^{1}$\thanks{E-mail: dfbu@shao.ac.cn}
Liang Chen,$^{1}$\thanks{E-mail: chenliang@shao.ac.cn}
Guobin Mou$^{2,3}$\thanks{E-mail: gbmou@whu.edu.cn}
Erlin Qiao$^{4,5}$
and Xiao-Hong Yang$^{6}$
\\
$^{1}$Shanghai Astronomical Observatory, Chinese Academy of Sciences, 80 Nandan Road, Shanghai 200030, China \\
$^{2}$School of Physics and Technology, Wuhan University, Wuhan 430072, China\\
$^{3}$WHU-NAOC Joint Center for Astronomy, Wuhan University, Wuhan 430072, China\\ 
$^{4}$Key Laboratory of Space Astronomy and Technology, National Astronomical Observatory, \\ Chinese Academy of Sciences, Beijing 100012, China\\ 
$^{5}$School of Astronomy and Space Sciences, University of Chinese Academy of Sciences, 19A Yuquan Road, Beijing 100049, China \\ 
$^6$Department of Physics, Chongqing University, Chongqing 400044, China
}

\date{Accepted XXX. Received YYY; in original form ZZZ}

\pubyear{2022}

\begin{document}
\label{firstpage}
\pagerange{\pageref{firstpage}--\pageref{lastpage}}
\maketitle

\begin{abstract}
Winds can be launched in tidal disruption event (TDE). It has been proposed that the winds can interact with the cloud surrounding the black hole, produce bow shocks, accelerate electrons, and produce radio emission. We restudy the wind-cloud interaction model. We employ the properties of winds found by the radiation hydrodynamic simulations of super-Eddington circularized accretion flow in TDEs. We can calculate the peak radio emission frequency, the luminosity at the peak frequency, and their time-evolution based on the TDEs wind-cloud interaction model. We find that the model predicted peak radio emission frequency, the luminosity at peak frequency, and their time evolution can be well consistent with those in TDEs AT2019dsg and ASASSN-14li. This indicates that in these two radio TDEs, the wind-cloud interaction mechanism may be responsible for the radio emission.
\end{abstract}




\section{Introduction}
Strong and transient radiation can be produced by tidal disruption events (TDEs). TDEs have been found in both soft X-ray bands (see \cite{Saxton2020} and \cite{Gezari2021} for reviews) and optical/UV band (see \cite{Velzen2020} and \cite{Gezari2021} for reviews). The soft X-rays are believed to be produced in the stellar debris accretion flow very close to the black hole (\cite{Metzger2016}; \cite{Strubbe2009}; \cite{Roth2016}; \cite{Dai2018}; \cite{Curd2019}). The origin of optical/UV emission of TDEs is still under debates. There are two competing models for the origins of optical/UV emission. In the first model, the optical/UV emission is directly generated in the process of shocks between the colliding debris streams (\cite{Piran2015}; \cite{Jiang2016}; \cite{Steinberg2022}). Alternatively, in the second model, the optically thick envelope/outflows surrounding the inner accretion flow reprocess the soft X-ray/EUV emission into optical/NUV bands (\cite{Strubbe2009}; \cite{Coughlin2014}; \cite{Roth2016}; \cite{Lodato2011}; \cite{Metzger2016};  \cite{Liu2021}; \cite{Bu2022}; \cite{Parkinson2022}; \cite{Wevers2022}; \cite{Thomsen2022}). \cite{Bu2022} find that the `circularized' super-Eddinton accretion flow in TDEs can produce strong outflows, which can reprocess the X-ray photons into optical/NUV bands. They also find that optical/UV radiation location, radiation temperature, luminosity as well as their evolutions are well consistent with observations.

A handful of TDE candidates are also found to have radio emission (see \cite{Alexander2020} for review). In the jetted TDEs (\cite{Bloom2011}; \cite{Burrows2011}; \cite{Zauderer2011}), the radio emission may be generated by the non-thermal electrons (NTe) accelerated in the shocks driven by jet (\cite{Giannios2011}). In addition to jet, numerical simulations show that large opening angle outflows can be generated in the stream-steam collision process (\cite{Jiang2016}; \cite{Lu2020}) or in the `circularized' super-Eddington accretion flow (\cite{Dai2018}; \cite{Curd2019}; \cite{Bu2022}; \cite{Thomsen2022}). In this paper, we define the large opening angle outflows as `TDE winds'. Observationally, the presence of TDE winds have been directly confirmed by the ultraviolet and X-ray spectra (\cite{Yang2017}; \cite{Kara2018}; \cite{Parkinson2020}). It is proposed that the TDE winds can interact with the circumnuclear medium (CNM) and the NTe accounting for the radio emission can be accelerated in the shocked CNM (\cite{Barniol2013}). The TDE wind-CNM model has been applied to some known radio TDEs (e.g. \cite{Matsumoto2021}; \cite{Matsumoto2022-1}).

In addition to hot diffuse CNM, there may exist clouds surrounding the supermassive black hole. \cite{Mou2022} (hereafter MOU22) proposed that the TDE wind-cloud interaction can produce bow shocks (see Figure \ref{fig:cartoon}), which can very efficiently convert wind kinetic energy into internal energy of shocked gas. A portion of the electrons become non-thermal relativistic electrons, which are responsible for the radio emission via synchrotron radiation. MOU22 found that the wind-cloud interaction model can roughly produce the observed evolution pattern of the peak frequency of the radio emission in several TDEs.

In MOU22, the properties of TDE winds (e.g., kinetic power, velocity, opening angle) are free parameters. Therefore, although the evolution pattern of the peak radio frequency can be consistent with observations, they can not give the radio luminosity (or spectrum) due to the lack of the exact properties of the TDE winds. \cite{Bu2022} (hereafter BU22) performed hydrodynamic simulations with radiative transfer to study the `circularized' accretion flow for TDEs. They found that strong TDE winds can be generated by the circularized super-Eddington accretion flow. In this paper, we re-study the TDEs wind-cloud interaction model. The properties of TDE winds are setting according to the simulation results in BU22. We calculate the evolution of the radio spectrum based on the winds-cloud interaction model and do comparisons to observations.

In Section 2, we introduce the properties of TDE winds and the wind-cloud interaction model. In section 3, we apply the model to several radio TDEs. We give a summary and discuss the results in Section 4.

\section{TDE wind-cloud interaction model}
\subsection{TDE winds}

In BU22, we have performed simulations with radiative transfer to study the circularized accretion flow/winds system in TDEs. In our simulations, we all assumed that the pericenter equals the tidal disruption radius. One of the most significant difference between simulations of the circularized TDE accretion flow in BU22 and others (\cite{Dai2018}; \cite{Curd2019}) is that we inject gas into our computational domain with the injection rate set according to the theoretically predicted debris gas fall back rate $\dot M_{\rm inject} \propto t^{-5/3}$. We find that in the presence of viscosity, the fallback debris can form a super-Eddington accretion flow with strong unbound winds (see Figure 2 in BU22).
In the simulations of BU22, the TDE winds are generated by the accretion flow inside two times the tidal disruption radius. The winds can move to significant large radius (outside the outer boundary of the simulation of $10^5 r_s$, with $r_s$ being the Schwarzschild radius).

\begin{figure}
\begin{center}
\includegraphics[width=0.45\textwidth]{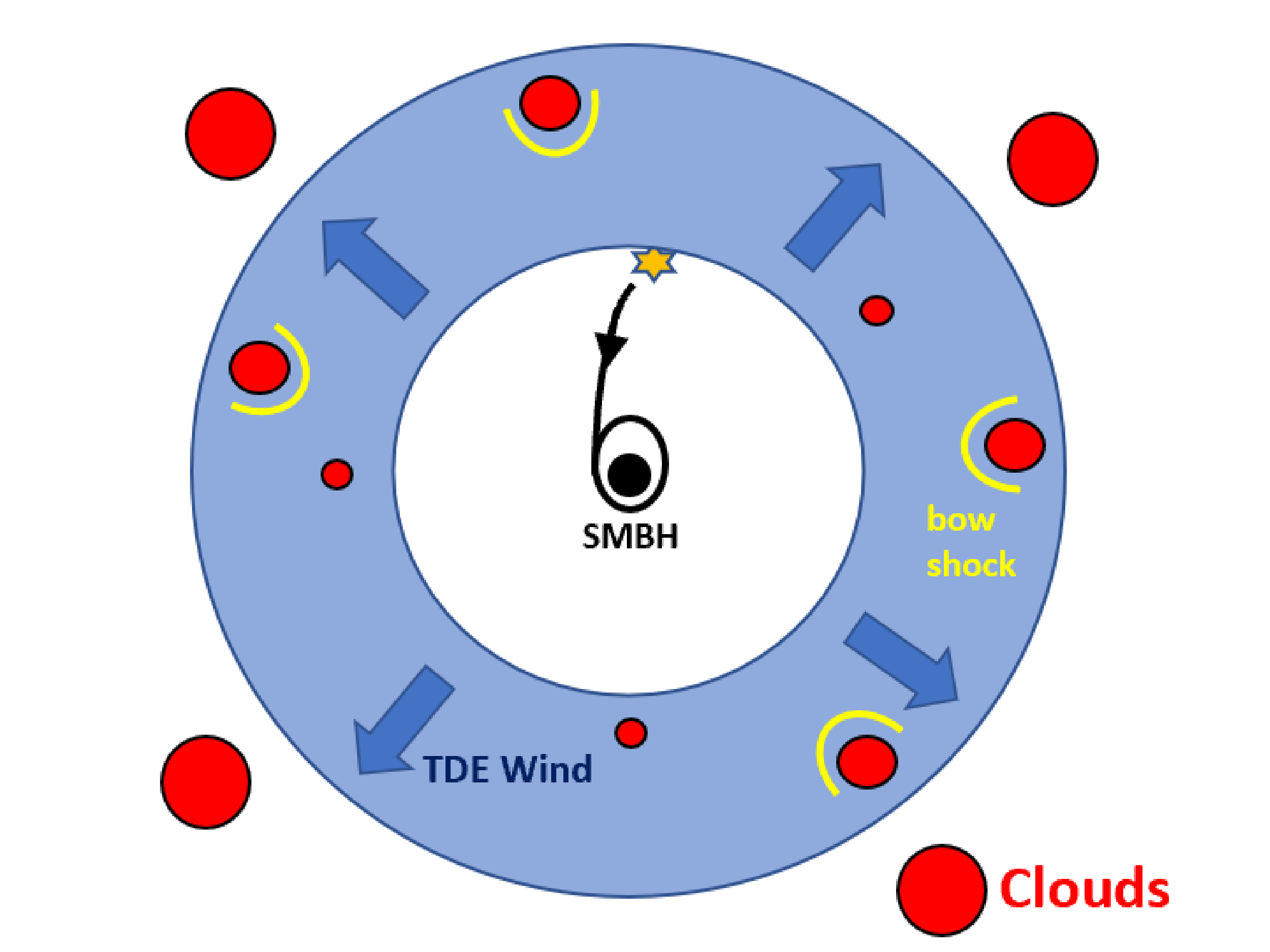}
\caption{Sketch of the TDE wind-cloud interaction model. The supermassive black hole (SMBH) is located at the center. The light blue shell marks the TDE winds. The red circles mark the dense clouds. The size of the clouds increases with the distance to the central black hole. The yellow arcs represent the bow shocks. In the model, the non-thermal electrons for the synchrotron emission are accelerated by bow shocks.}
\label{fig:cartoon}
\end{center}
\end{figure}

\begin{figure*}
\begin{center}
\includegraphics[width=0.45\textwidth]{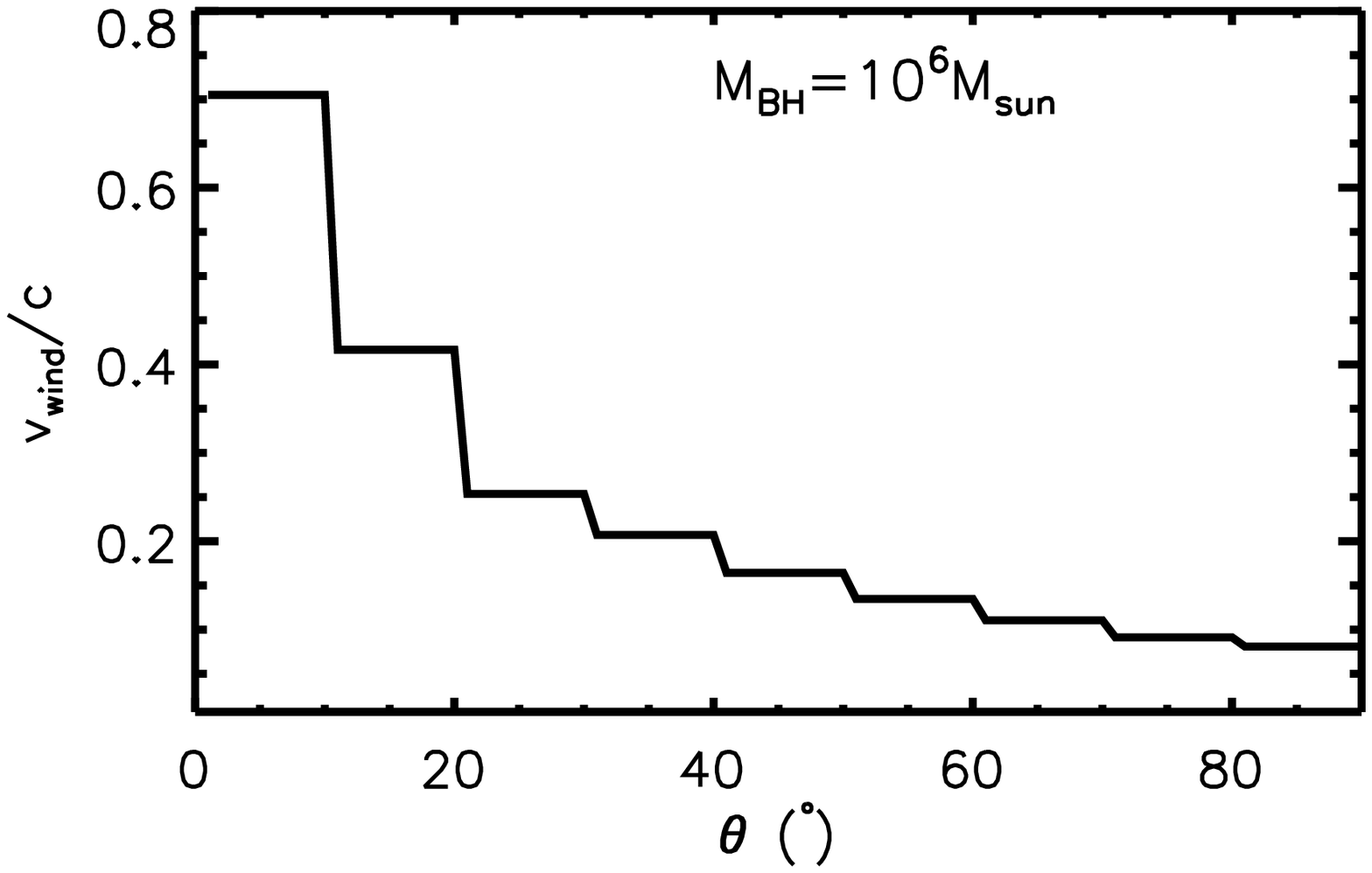}
\includegraphics[width=0.45\textwidth]{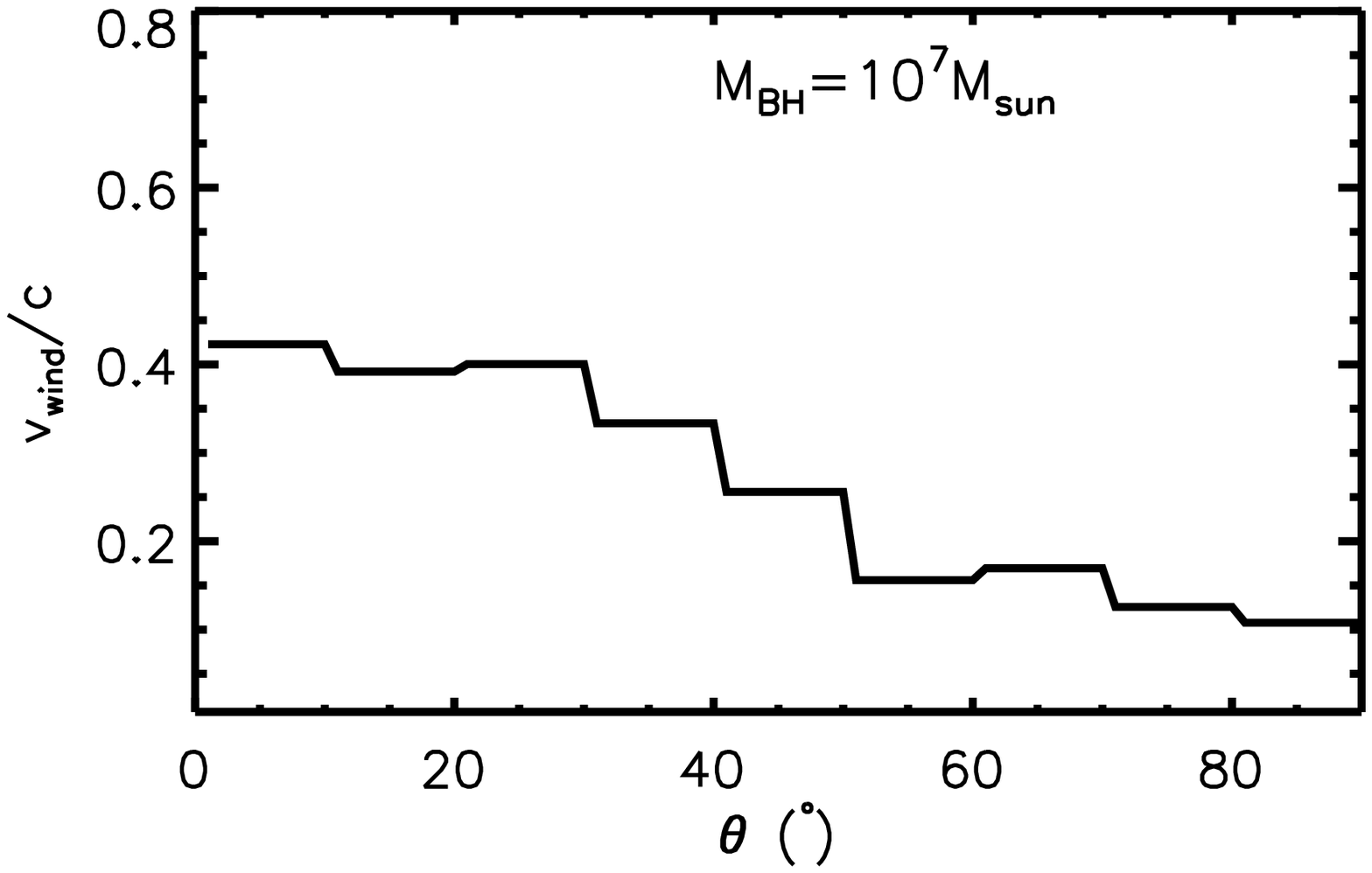}\\
\includegraphics[width=0.45\textwidth]{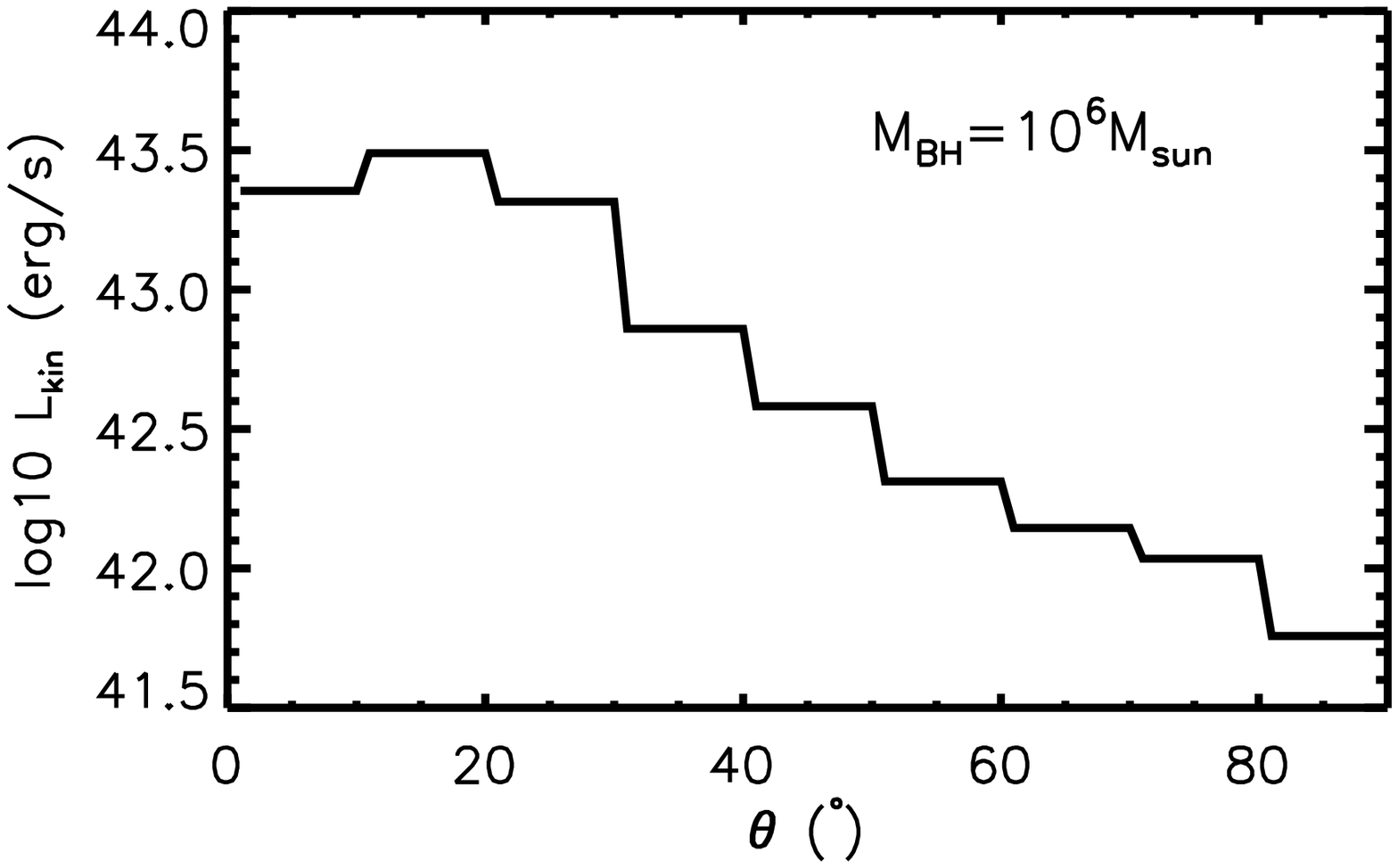}
\includegraphics[width=0.45\textwidth]{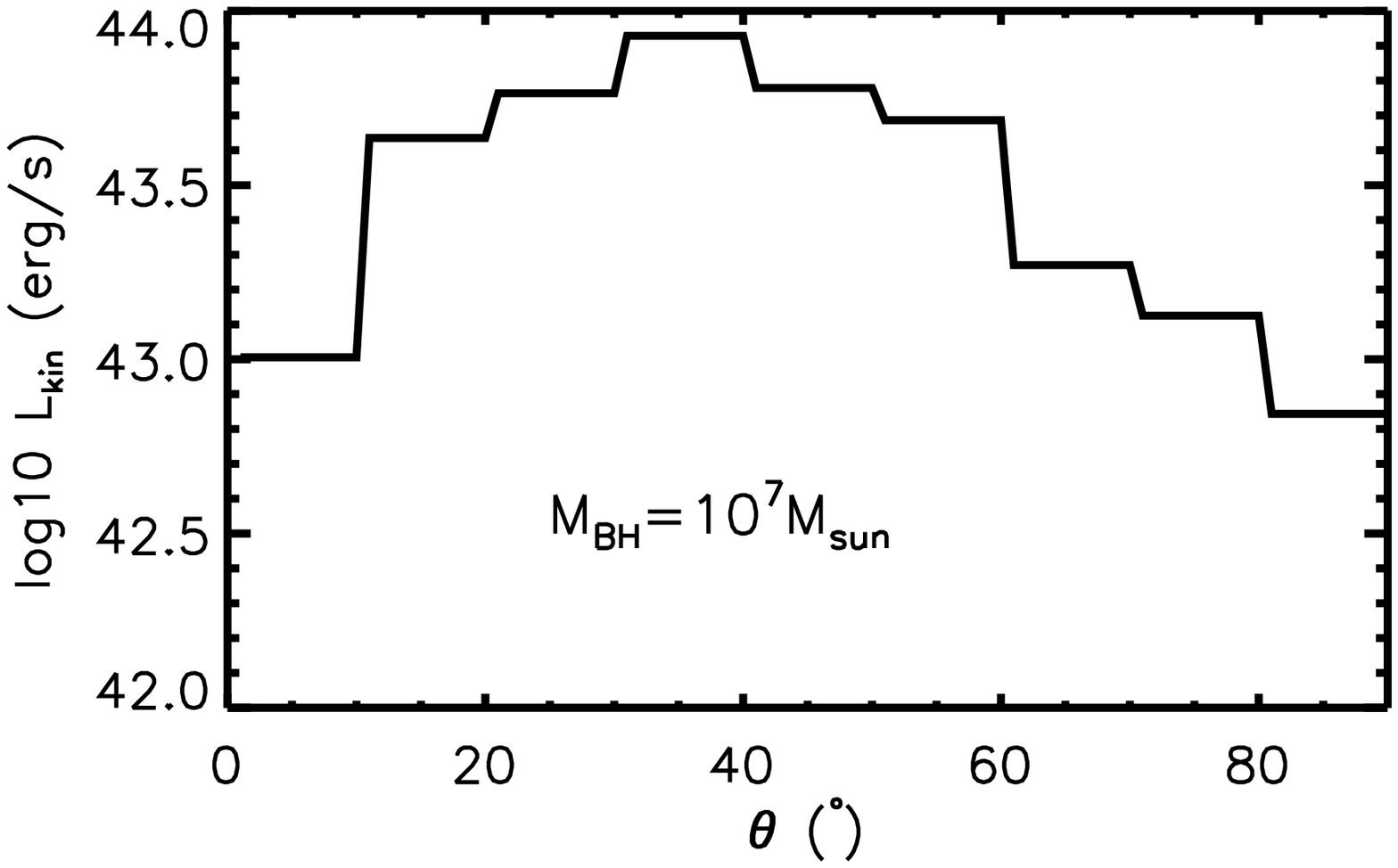}\\
\includegraphics[width=0.45\textwidth]{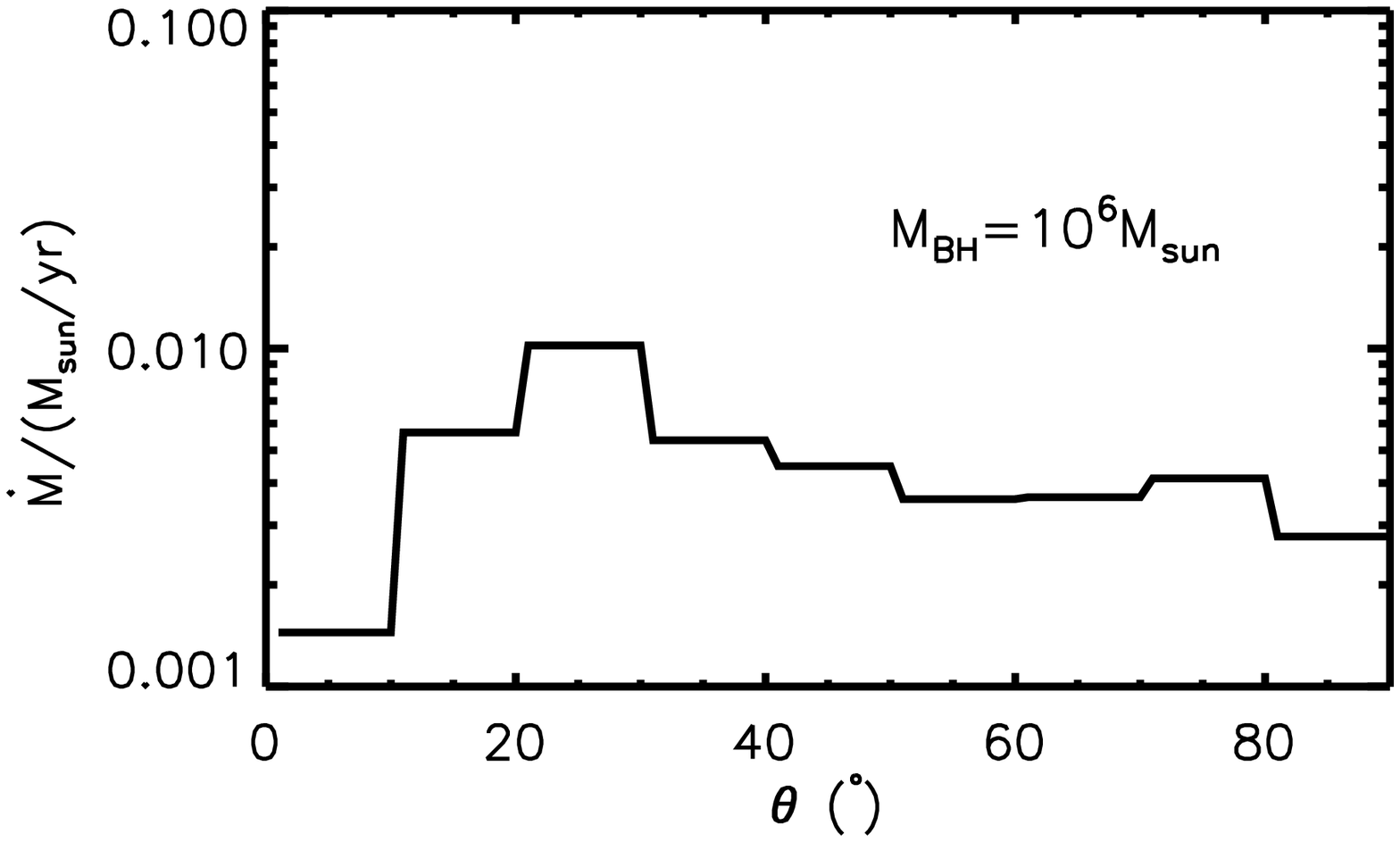}
\includegraphics[width=0.45\textwidth]{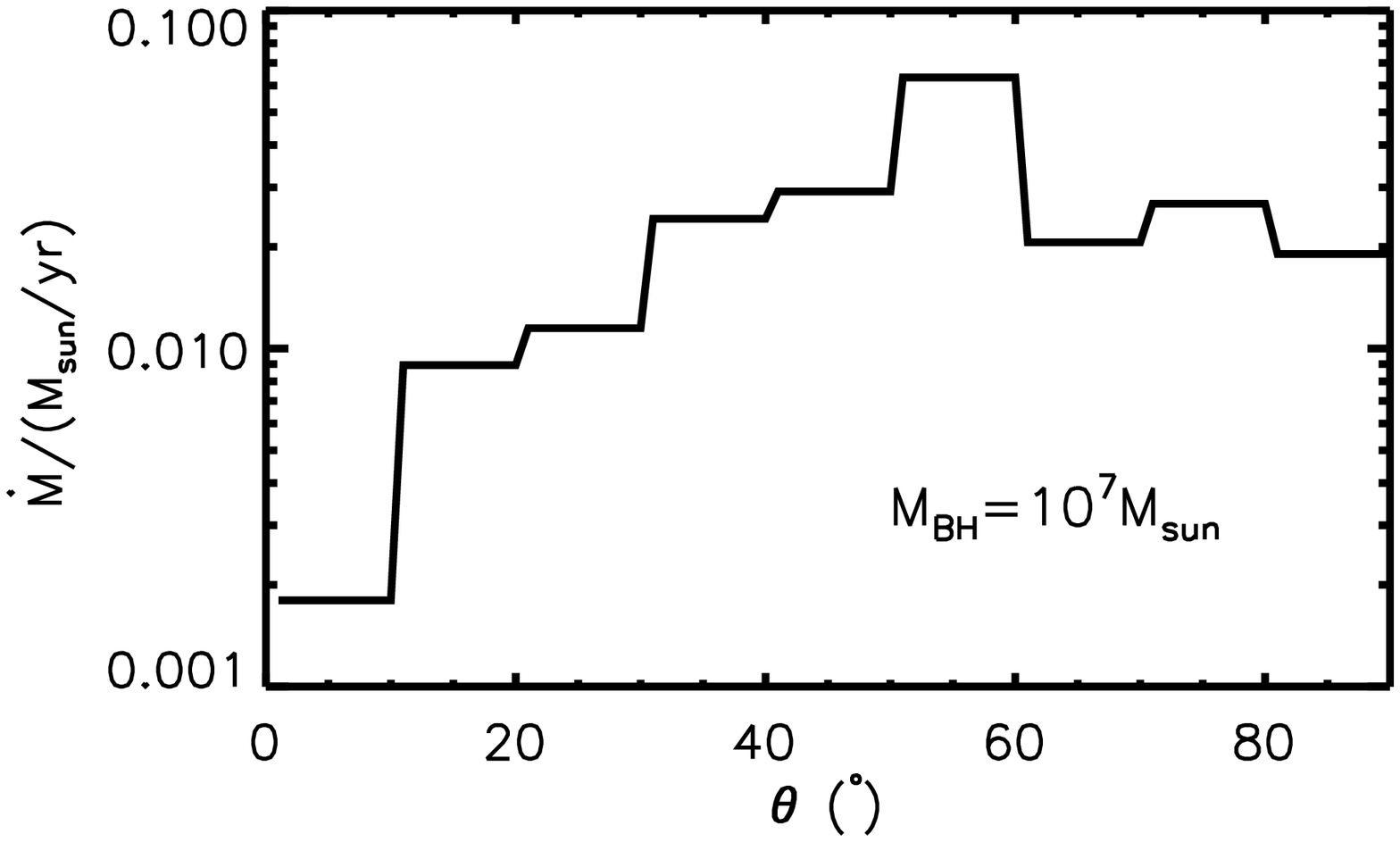}
\caption{Velocity (top panels), kinetic power (middle panels) and mass flux (bottom panels) of wind head for model M6 with $M_{\rm BH}=10^6 M_\odot$ (left panels) and model M7 with $M_{\rm BH}=10^7 M_\odot$ (right panels).}
\label{fig:wind}
\end{center}
\end{figure*}

In our radio emission model, we assume that the TDE winds head collides with condensed clouds and produces the bow shocks which accelerates the power-law NTe (see Section 2.2). The luminosity and peak radiation frequency of the radio emission depends on the properties (kinetic power, winds velocity, opening angle) of TDE winds head. We introduce the properties of winds head (see Appendix A for the method to find the wind head). The simulation of BU22 has a polar computational domain from the north rotational axis to the midplane ($0^\circ \leq \theta \leq 90^\circ$). They assume that the flows above and below the midplane are symmetric with respect to the midplane. BU22 showed that the winds are present in the region from the accretion flow north rotational axis to the midplane ($0^\circ \leq \theta \leq 90^\circ$). Generally, the velocity of the winds decreases from the rotational axis to the midplane. We cut the polar computational domain of BU22 into 9 equally spaced bins. Each bin has a polar angle interval of 10 degrees. For each bins, we calculate the power of winds head as,
\begin{equation}
L_{\rm kin} = 2\pi r_{\rm head}^2 \int_{\theta_{\rm bottom}}^{\theta_{\rm up}} \rho_{\rm wind} \frac{1}{2} v_{\rm wind}^3 \sin{\theta}d\theta
\end{equation}
where, $\theta_{\rm bottom}$ and $\theta_{\rm bottom}$ are the lower and upper boundary of each polar bin. $r_{\rm head}$ is the location of the winds head.
The mass flux of the winds head for each bin is,
\begin{equation}
\dot M = 2\pi r_{\rm head}^2 \int_{\theta_{\rm bottom}}^{\theta_{\rm up}} \rho_{\rm wind} v_{\rm wind} \sin{\theta}d\theta
\end{equation}
Then, the velocity of the wind head for each bin is calculated as
\begin{equation}
v_{\rm wind} = \sqrt{\frac{2 L_{\rm kin}}{\dot M}}
\end{equation}
The wind head moves from small to large radii, therefore, in Equations (1) and (2), the value of $r_{\rm head}$ increases with time. However, we note that the TDE winds is unbound (see Figure 2 in BU22), with the velocity significantly larger than the local escape velocity. In such a sense, the effect of gravitational pull on winds by the black hole is negligible. The velocity of the winds is almost non-variable with the outwards motions. We can safely assume that when the wind moves from small to large radii, the velocity does not change. The power of winds head is also almost non-variable. Therefore, we can calculate the power and velocity of the winds head based on a specific time snapshot data.
The simulations of BU22 have a starting time which corresponds to the time at which the maximum debris fallback rate is achieved. We can roughly assume that the starting time of the simulations corresponds to the outburst time of TDEs. In BU22, we have two simulations with the black hole mass $M_{\rm BH}$ corresponding to $10^6 M_\odot$ and $10^7 M_\odot$, respectively. For the model M6 with $M_{\rm BH}=10^6 M_\odot$, we calculate the properties of wind head at $t=30$ days after the outburst. For the model M7 with $M_{\rm BH}=10^7 M_\odot$, we calculate the properties of wind head at $t=197$ days after the outburst.

Figure \ref{fig:wind} shows the velocity (top panels), kinetic power (middle panels) and mass flux (bottom panels) of wind head  for each bins for model M6 with $M_{\rm BH}=10^6 M_\odot$ (left panels) and model M7 with $M_{\rm BH}=10^7 M_\odot$ (right panels). It can be seen that the wind head velocity generally decreases from the rotational axis towards the midplane. The maximum velocity of wind head is $\sim 0.7 c$; the minimum velocity of wind head is $\sim 0.1 c$. The kinetic power of wind head for some bin can be as high as $10^{44}$ erg/s. The mass flux is non-uniformly distributed with $\theta$. The mass flux of wind has its lowest value close to the rotational axis due to the low density there. The maximum value of mass flux is in the region of $20^\circ - 30^\circ$ for model M6 and $50^\circ - 60^\circ$ for model M7.  We note that the total kinetic power and mass flux of the wind head is 2 times the value of the sum of the powers of all bins. In BU22, we only simulate the region above the midplane. Actually, the region below the midplane has the same amount of kinetic power and mass flux. Therefore, there is a factor of 2 when we calculate the total kinetic power and mass flux. The total kinetic power in model M6 is $1.8 \times 10^{44}$ erg/s. For model M7, the total kinetic power is $6.9 \times 10^{44}$ erg/s. We note that in \cite{Curd2019}, they find that for a $10^6M_\odot$ black hole, the kinetic power of wind generated from a weakly magnetized accretion flow is roughly $4 \times 10^{44}$ erg/s, which is very close to the values obtained in BU22. The mass flux of wind in model M6 is $8.7 \times 10^{-2} M_\odot \ {\rm yr}^{-1}$. For model M7, the mass flux of wind is $4.2 \times 10^{-1} M_\odot \ {\rm yr}^{-1}$.

\subsection{Non-thermal electrons}
When the TDE winds move outwards and interact with condensed clouds, bow shocks will be generated, which propagate inwards and shock the winds (see Figure \ref{fig:cartoon}). A fraction ($\epsilon_s$) of the winds kinetic energy is converted into internal energy of shocked gas. $\epsilon_s$ is roughly equal to the covering factor of the clouds $c_f$. A fraction ($\epsilon_e$) of the internal energy of shocked gas can be converted to NTe. In the present paper, we set $\epsilon_e = 0.1$.

The NTe distribution has a power law form of energy ${\rm d} N(\gamma_e)= A\gamma_e^{-p} {\rm d} \gamma_e $, with $\gamma_e$ being the Lorentz factor (LF) of electrons. The minimum value of the electron Lorentz factor ($\gamma_m$) depends on the bulk motion velocity of TDE winds (\cite{Barniol2013}). There is a critical velocity of wind, $v_{\rm DN} = c\sqrt{16m_{\rm e}/(m_{\rm p}\bar \epsilon_{e})}$ (\cite{Sironi2013}), with $\bar{\epsilon}_e = 4 \epsilon_e (p-2)/(p-1)$ and $c$ being speed of light. When the velocity of winds $v_{\rm wind} < v_{\rm DN}$, it is in deep-Newtonian phase. For deep-Newtonian case, $\gamma_m = 2$ (\cite{Huang2003}) and only a fraction $(v_{\rm wind}/v_{\rm DN})^2$ of electrons in post-shock winds is in the power-law distribution with $\gamma_e \geq 2$. For $v_{\rm wind} > v_{\rm DN}$, it is in the mildly relativistic regime and $\gamma_m \sim 230 \bar{\epsilon}_e (v_{\rm wind}/c)^2 > 2$. In the mildly relativistic regime, all of the electrons in post-shock winds participate in the pow-law distribution with $\gamma_m \geq 2$. For the typical values adopt in this paper $\bar\epsilon_e = 0.1$, we have $v_{\rm DN}=0.3c$. The power law index $p$ is in the range $2.5-3.0$ for deep-Newtonian case (e.g., \cite{Chevalier2006}) and in the range $2.1-2.5$ for the mildly relativistic case.

\cite{Mou2021} showed that for wind-cloud interaction, the high pressure behind the bow shock can push the shocked winds to expand towards the back of the cloud. In this process, the energy of the NTe will decrease due to adiabatic loss. The adiabatic cooling time-scale of NTe is proportional to the time needed for the wind head to cross the cloud $t_{\rm ad} = k_{\rm bow} R_c/v_{\rm wind}$, with $R_c$ being the cloud radius and $k_{\rm bow} \sim 10$ (\cite{Mou2021}). The radio emission for TDEs can last for several hundred days after the peaks luminosity at other bands (\cite{Alexander2020}). If we assume that the winds can be generated at the beginning of the disruption, for a typical wind velocity of several tenth of speed of light (see next subsection), the winds can move to roughly $10^{18} {\rm cm}$ after several hundred days. In AGNs, the typical cloud size $R_c = \eta r$, with $\eta$ being of the order of $10^{-3}-10^{-2}$ and $r$ being the distance to the central supermassive black hole (\cite{Netzer2015}). If we take $\eta=10^{-3}$ as a fiducial value for all the galaxies, at the location $r=10^{18} {\rm cm}$, $t_{\rm ad} \sim k_{\rm bow} 10^{-3} r / v_{\rm wind} \sim 10^6 {\rm s} \sim 10 {\rm days}$ for a wind velocity of $v_{\rm wind} = 0.5 c$. \footnote{If the wind speed is constant, the radius is given by $r=v_{\rm wind} t$. We have $t_{\rm ad} = k_{\rm bow} \eta t \sim {\rm days} (k_{\rm bow}/10)(\eta/10^{-3})(t/100{\rm days})$, which does not depend on the velocity of wind.} In the region $r < 10^{18}{\rm cm}$, $t_{\rm ad}$ will be smaller. For a black hole of mass of $10^6-10^7 M_\odot$ (with $M_\odot$ being solar mass) disrupting a solar type star, the debris fall back rate can stay above the Eddington accretion rate for several hundred days (\cite{Roth2020}). If the black hole mass is close to $10^6 M_\odot$, the super-Eddington phase can be roughly 2 years (\cite{Rees1988}; \cite{Lodato2009}; \cite{Wu2018}; \cite{Roth2020}). For such a high gas fall back rate, the accretion flow onto the central black hole can be super-Eddington and radiation pressure dominated. Such a super-Eddington accretion flow can drive strong winds (\cite{Curd2019}; BU22). The duration of launching energetic winds $t_{\rm bst}$ is of the order of hundreds of days. $t_{\rm bst}$ is orders of magnitude longer than $t_{\rm ad}$. When the head of winds sweeps the cloud, the clouds can be strongly compressed, which decreases the covering factor of the clouds significantly. A significantly reduced covering factor makes the shock energy reduced significantly. In such a sense, the bow shock at a distance $r$ can be regarded as a transient. Given that $t_{\rm bst} >> t_{\rm ad}$, bow shocks at $r$ can last for a timescale comparable to $t_{\rm ad}$. In this case, the NTe energy of the bow shocks at distance $r$ is $\sim c_f \epsilon_e L_{\rm kin} t_{\rm ad}$ (MOU22), with $L_{\rm kin}$ being the kinetic power of winds at winds head. The transient NTe at distance $r$ will generate a transient radio emission. When the winds head moves to a larger radius, new bow shocks and NTe will be generated. The location of radio emission moves to the larger radius. In our model, the distance of the radio emission region to the central black hole increases with time. The energy of NTe with $\gamma_{\rm e} > \gamma_{\rm m}$ at radius $r$ is,
\begin{equation}
E_{\rm nte}(r) = c_f \epsilon_e L_{\rm kin} t_{\rm ad} (r) = N_{\rm e} (r) m_{\rm e} c^2 \gamma_{\rm m} \frac{p-1}{p-2}
\end{equation}
Where $N_{\rm e} (r)$ is the total number of NTe and $m_{\rm e}$ is the electron mass. Note that $L_{\rm kin}$ is the kinetic power of winds at winds head.

\subsection{Radio spectrum via synchrotron emission}
The magnetic field can be amplified in the bow shock (\cite{Bell2001}; \cite{Schure2012}). We can parameterize the magnetic pressure as $B^2/8\pi=\epsilon_{\rm B} \rho_{\rm wind} v^2_{\rm wind}$. The value of $\epsilon_{\rm B} \sim 10^{-2}-0.1$ in the shock downstream (\cite{Volk2005}). Adopting $r=v_{\rm wind}t$, we have
\begin{equation}
B = (8\pi\epsilon_{\rm B} \rho_{\rm wind} v^2_{\rm wind})^{1/2} = 46 \ {\rm G} \ \epsilon_{\rm B, -1}^{1/2} \dot {m}^{1/2} \Omega^{-1/2} v_{9}^{-1/2} t_2^{-1}
\end{equation}
In above equation, $\epsilon_{\rm B, -1} = \epsilon_{\rm B}/0.1$,
the mass flux of wind is scaled as $ \dot {m} = \dot {M}/1 M_\odot {\rm yr}^{-1}$ (see Equation 2 for calculation of wind mass flux). $\Omega$ is the opening solid angle of wind, velocity of wind is scaled as $v_9 = v_{\rm wind}/10^9 {\rm cm \ s^{-1}}$, time is scaled as $t_2 = t / 10^2 {\rm day}$. Note that the values of $\dot m$, $\Omega$ and $v_9$ are different from bin to bin (see Figure \ref{fig:wind}).

For an electron with LF $\gamma_{\rm m}$, the typical synchrotron emission frequency is
\begin{equation}
\nu_{\rm m} = \frac{\gamma^2_{\rm m} e B}{2 \pi m_{\rm e} c} = 1.3 \times 10^8 {\rm Hz} \gamma^2_{\rm m} \epsilon_{\rm B, -1}^{1/2} \left( \frac{\dot m}{\Omega v_9} \right)^{1/2} t_2^{-1}
\end{equation}
where $e$ is the electron charge. The luminosity emitted by one electron  at frequency $\nu_{\rm m}$ is approximately $\nu_{\rm m} P_{\nu_{\rm m}} \sim (4/3)\sigma_{\rm T} c \gamma^2_{\rm m} (B^2/8\pi)$, with $\sigma_{\rm T}$ being the Thomson cross-section. The luminosity at frequency $\nu_{\rm m}$ is
\begin{equation}
L_{\nu_{\rm m}} \simeq N_{\rm e} \frac{1}{\nu_{\rm m}}\frac{4}{3}\sigma_{\rm T}c\gamma^2_{\rm m}\frac{B^2}{8\pi}
\end{equation}
Combing Equations (4), (5), and (6), we have
\begin{equation}
\begin{aligned}
L_{\nu_{\rm m}} \simeq & \frac{\pi^{\frac{1}{2}}\sigma_{\rm T}}{3\gamma_{\rm m} e} k_{\rm bow} c_f \epsilon^{\frac{1}{2}}_{\rm B} \bar{\epsilon}_{\rm e} L^{\frac{3}{2}}_{\rm kin} (\Omega \ v^3_{\rm wind})^{-\frac{1}{2}} \eta \\  = & 1.5 \times 10^{35} {\rm erg \ s^{-1}Hz^{-1}} \gamma_{\rm m}^{-1} k_{\rm bow} c_f \\
& \times \epsilon_{\rm B, -1}^{1/2} \bar\epsilon_{\rm e, -1} \dot{m}^{3/2}\Omega^{-1/2} v_{9}^{3/2} \eta
\end{aligned}
\end{equation}
where $\bar\epsilon_{\rm e, -1} = \bar\epsilon_{\rm e}/0.1 $ and the cloud size is scaled as $R_{\rm c, 13} = R_c/10^{13} {\rm cm}$.

The peak value of $L_\nu$ is determined by the synchrotron self-absorption (SSA). The self-absorption frequency $\nu_{\rm a}$ is derived from the condition that the optical depth of the emission region $\alpha_\nu l=1$, where $\alpha_\nu \simeq (\pi^{3/2}/4)3^{(p+1)/2}AeB^{-1}\gamma_{\rm m}^{-p-4}\nu_{\rm m}^{(p+4)/2}\nu^{-(p+4)/2}$ is the absorption coefficient and $l$ is the size of the emitting region. We assume that $l= R_{\rm c}$. Note here that $A$ is the coefficient of the power law distribution of electrons in unit volume. As introduced above, in the deep-Newtonian phase, only a fraction $(v_{\rm wind}/0.3c)^2$ of the electrons in post-shock winds are in power-law distribution with $\gamma_{\rm e} \geq 2$. In the mildly relativistic phase, all the electrons in post-shock winds are in power-law distribution Therefore, we have
\begin{equation}
A=(p-1)\gamma_{\rm m}^{p-1} \rho_{\rm wind} (r)/m_{\rm p} \cdot \Lambda
\end{equation}
where $\Lambda \equiv \min[(v_{\rm wind}/0.3c)^2,1]$ and $m_{\rm p}$ being the proton mass.
The value of $\nu_{\rm a}$ is
\begin{equation}
\begin{aligned}
\nu_{\rm a}  = \nu_{\rm m} \left( \frac{\pi 3^{(p+1)/2} (p-1) e  \Lambda L_{\rm kin}^{1/2}} {8 m_{\rm p} \epsilon_{\rm B}^{1/2} \gamma_{\rm m}^{5} v_{\rm wind}^{5/2} \Omega^{1/2}} \frac{R_{\rm c}}{r} \right)^{2/(p+4)} \\
\end{aligned}
\end{equation}

\begin{align}
=\begin{cases}
1.6\times 10^8 {\rm Hz} f({\rm p}) f_0({\rm p}) \epsilon_{\rm B, -1}^{({\rm p+2})/2({\rm p+4})} \bar\epsilon_{\rm e, -1}^{2/({\rm p+4})} \\ \times \left( \frac{\dot{m}}{\Omega} \right)^{({\rm p+6})/2({\rm p+4})} f_1({\rm p}) \eta^{2/({\rm p+4})} v_9^{-(({\rm p+2})/2({\rm p+4}))} t_{2}^{-1}  \\ ~~~~~~~~ (v_{\rm wind} \leqslant v_{\rm DN}) ~~~ \\
3.9\times 10^9 {\rm Hz} f_2({\rm p}) f_0({\rm p}) \epsilon_{\rm B, -1}^{({\rm p+2})/2({\rm p+4})} \bar\epsilon_{\rm e, -1}^{2({\rm p-1})/({\rm p+4})} \\ \times \left( \frac{\dot{m}}{\Omega} \right)^{({\rm p+6})/2({\rm p+4})} \beta^{({\rm 7p-22})/2({\rm p+4})} f_3({\rm p}) \eta^{2/({\rm p+4})} v_9^{2/({\rm p+4})} t_{2}^{-1}  \\ ~~~~~~~~ (v_{\rm wind} > v_{\rm DN}) ~~~ \\
\end{cases}
\end{align}
where $\beta={v_{\rm wind}/c}$. The parameters $f_0({\rm p}) = 10^{({\rm p+2})/2({\rm p+4})}$, $f(\rm p) = [2.6 \times 10^{16} \times 3^{\rm p} ({\rm p-1})^2]^{1/({\rm p+4})}$, $f_1({\rm p}) = (864.5)^{2/(p+4)}$, $f_2(\rm p) = [160 \times 3^{\rm p} ({\rm p-1})^2]^{1/({\rm p+4})}$ and $f_3({\rm p}) = (8.645)^{2/(p+4)}$ for cgs units. As ${p}$ increases from 2.1 to 3.6, $f_0({\rm p})$ increases monotonically from 2.17 to 2.34, $f_1({\rm p})$ decreases monotonically from 9.18 to 5.93, $f({\rm p})$ decreases monotonically from 739 to 313, $f_2({\rm p})$ increases monotonically from 3.46 to 4.22, $f_3({\rm p})$ decreases monotonically from 2.03 to 1.76 (MOU22).

For $\nu_{\rm a} > \nu_{\rm m}$, we have the peak synchrotron luminosity as,
\begin{equation}
L_{\nu_{\rm a}} = L_{\nu_{\rm m}} (\nu_{\rm a}/\nu_{\rm m})^{(1-p)/2}
\end{equation}

\begin{align}
=\begin{cases}
2.7\times 10^{31} {\rm erg \ s^{-1}} {\rm Hz}^{-1} f({\rm p})^{\rm (1-p)/2} k_{\rm bow} c_f \\ \times f_4({\rm p}) \epsilon_{\rm B, -1}^{({\rm 2p+3})/2({\rm p+4})} \bar\epsilon_{\rm e, -1}^{5/({\rm p+4})} {\dot{m}}^{({\rm 2p+13})/2({\rm p+4})} \Omega^{-5/(2({\rm p+4}))}\\f_5({\rm p})\eta^{5/({\rm p+4})}  v_9^{({\rm 2p+13})/2({\rm p+4})}  \\ ~~~~~~~~ (v_{\rm wind} \leqslant v_{\rm DN}) ~~~ \\
1.3\times 10^{31} {\rm erg \ s^{-1}} {\rm Hz}^{-1} f_2({\rm p})^{\rm (1-p)/2} k_{\rm bow} c_f \\ \times f_4({\rm p}) \epsilon_{\rm B, -1}^{({\rm 2p+3})/2({\rm p+4})} \bar\epsilon_{\rm e, -1}^{5({\rm p-1})/({\rm p+4})} {\dot{m}}^{({\rm 2p+13})/2({\rm p+4})} \Omega^{-5/(2({\rm p+4}))}\\ \beta^{({\rm 22p-37})/2({\rm p+4})} f_5({\rm p}) \eta^{5/({\rm p+4})} v_{9}^{5/({\rm p+4})}  \\ ~~~~~~~~ (v_{\rm wind} > v_{\rm DN}) ~~~ \\
\end{cases}
\end{align}
The parameters $f_4({\rm p}) = 10^{({\rm 2p+3})/2({\rm p+4})}$, $f_5({\rm p}) = (864.5)^{5/(p+4)}$ for cgs units. As ${p}$ increases from 2.1 to 3.6, $f_4({\rm p})$ increases monotonically from 3.89 to 4.69, $f_5({\rm p})$ decreases monotonically from 255 to 86.

The spectrum of the synchrotron emission taken into account SSA is
\begin{align}
L_{\nu}=\begin{cases}
L_{\nu_{\rm a}} (\nu/\nu_{\rm a})^{5/2}     ~~~~~~~~~~~~~~ (\nu < \nu_{\rm a}) ~~~ \\
L_{\nu_{\rm a}} ({\nu/\nu_{\rm a}})^{(1-p)/2}  ~~~~~~~~~ ( \nu \geqslant \nu_{\rm a})
\end{cases}
\end{align}

As introduced in section 2.1, we cut the TDE winds into 9 angular bins. When we calculate the radio emission, each angular bin is independent. We calculate the radio emission from the interaction of each bin with the clouds. Then, we add the contributions from each bin to obtain the total radio emission.

\section{Applications}
Given that the properties of TDE winds are known, the radio emission peak frequency ($\nu_{\rm a}$) and the peak value of $L_{\nu_{\rm a}}$ depends on parameters $k_{\rm bow}$, $c_f$, $\epsilon_{\rm B}$, $\epsilon_{\rm e}$, the ratio of $R_{\rm c}/r$, and the power law index of NTe $p$.

\begin{figure*}
\begin{center}
\includegraphics[width=0.45\textwidth]{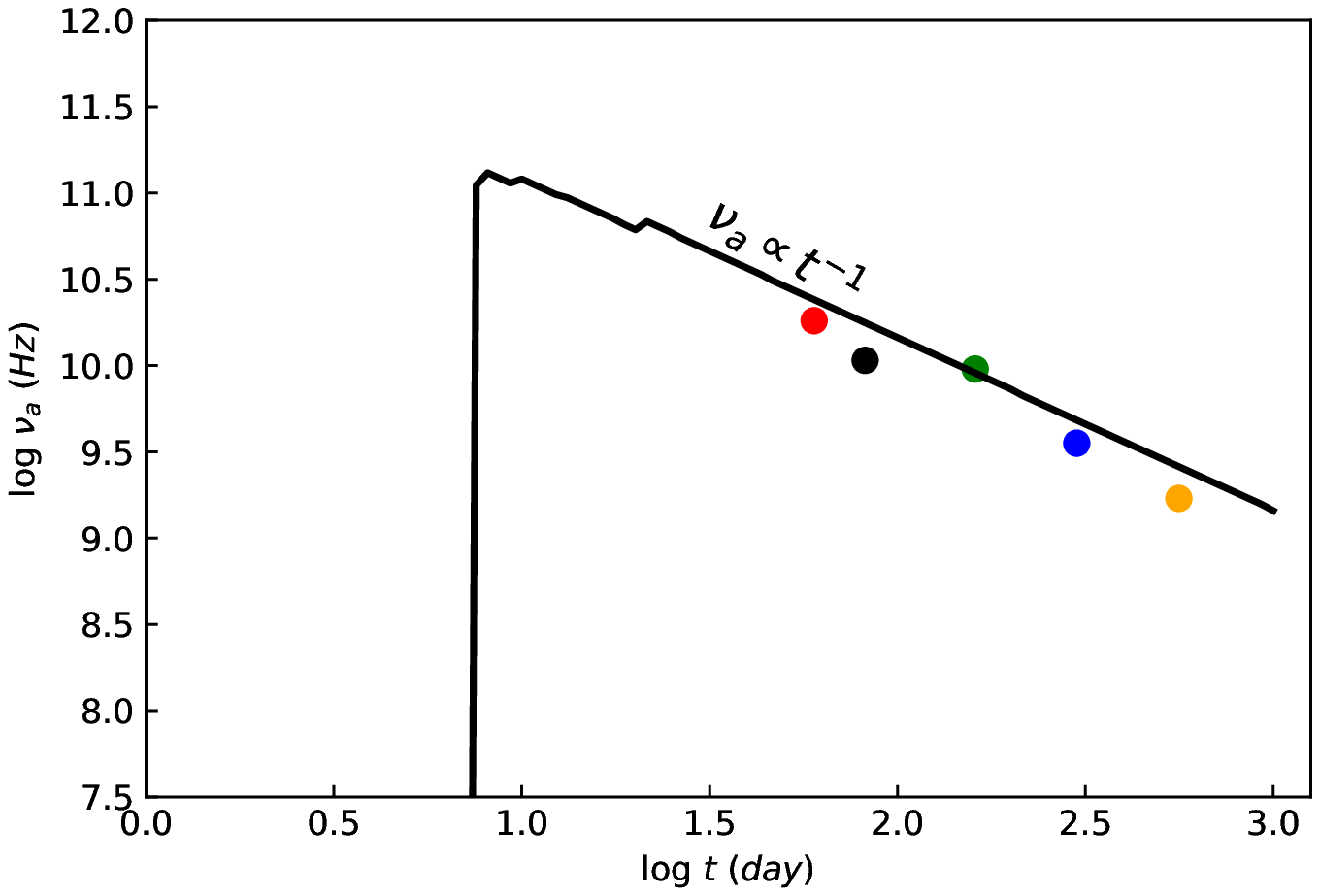}
\includegraphics[width=0.45\textwidth]{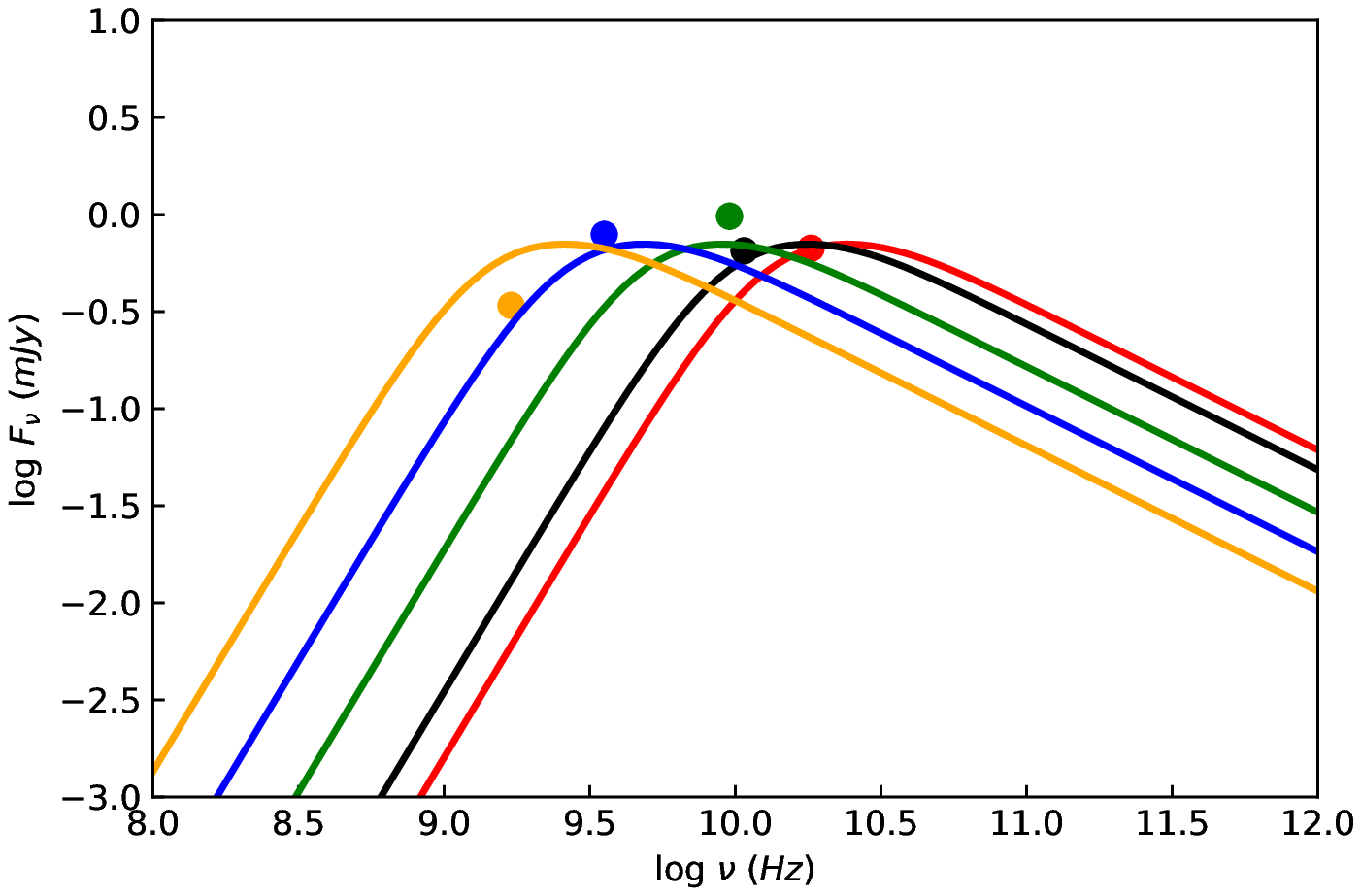}\\
\caption{Time evolution of the peak frequencies (left panel) and SED (right panel). In both panels, the colored dots are for AT2019dsg. The dots colored red, black, green, blue, and yellow correspond to 60 day, 82 day, 161 day, 300 day and 561 day after the outburst date. In the left panel, the solid line is the wind-cloud interaction model predicted evolution of the peak frequency. Note that the peak frequency shown in the left panel is equivalent to the SSA frequency. In the right panel, the colored solid lines are the model predicted SEDs at different snapshot after outburst. The colored dots in the right panel show the peak flux density $F_{\nu, \rm a}$. The details for the parameters of the model are introduced in Section 3.1.}
\label{fig:SED7}
\end{center}
\end{figure*}

We apply the wind-cloud interaction model to some radio TDEs. We exclude jet-induced radio TDEs. We also choose radio TDEs with optical or X-ray outbursts and peak frequencies detections. Four TDE candidates with radio flares occurring a few months after outbursts are obtained. They are AT2019dsg (\cite{Cendes2021}), ASASSN-14li (\cite{Alexander2016}), CSS161010 (\cite{Coppejans2020}) and ASASSN-15oi (\cite{Horesh2021}).

\subsection{AT2019dsg}
AT2019dsg was discovered on 2019 April 9 and classified as a TDE based on its optical spectrum (\cite{van2021}). This source has a redshift $z=0.051$. The mass of the black hole is $\log10 (M_{\rm BH}/M_\odot) = 6.7 \pm 0.4 $ (\cite{Cannizzaro2021}). Multi-frequency radio observations of AT2019dsg on a timescale of 55 to 560 days after disruption were obtained (\cite{Cendes2021}). The peak radio frequency ($\nu_{\rm a}$), the peak flux density ($F_{\nu, \rm a}$), and their time-evolution are obtained from the SED fitting of observations at successive epochs.

The black hole mass of AT2019dsg is close to $10^7 M_\odot$. When we calculate the radio emission based on the TDE wind-cloud interaction model, we use the wind properties of model M7 (see the bottom panels of Figure \ref{fig:wind}). We adopt parameters $c_f = 0.2$, $\epsilon_{\rm e} = \epsilon_{\rm B} =0.1$, $p=2.5$. Following MOU22, we assume that the cloud size $R_{\rm c}=\eta r$, with $\eta = 10^{-3}$. The typical broad line region has a distance to the central black hole of $1-10$ light-days (\cite{Netzer2015}). We assume that the clouds are distributed outside $10^{1/2}$ light-days away from the black hole.

Figure \ref{fig:SED7} shows the time evolution of the peak frequency $\nu_{\rm a}$ (left panel), the peak flux density $F_{\nu, \rm a}$ and SED (right panel). Note that the peak frequency is equivalent to the SSA frequency. The solid line in the left panel is the model predicted evolution of the peak frequency. The fastest winds have a velocity of $\sim 0.42 c$ (see the bottom left panel of Figure \ref{fig:wind}). We assume that the clouds are distributed outside $10^{1/2}$ light-days. It takes $\sim 7.5$ days for the TDE winds to arrive at a distance of $10^{1/2}$ light-days. Then the interaction of winds and clouds can produce radio emission. Therefore, the radio emission arises after 7.5 days after the outburst as shown by the solid line in the left panel of Figure \ref{fig:SED7}. Note that the inner boundary of the distribution of the clouds is very arbitrary. Thus, the time delay between the radio emission and the outburst of the TDEs radiation at other bands can not be taken seriously. However, we note that the evolution pattern of the peak frequency and the peak flux density are not affected by the arbitrary choice of the innermost location of the clouds. In other words, the absolute values of the peak frequency and the peak flux density at any specific time are not affected by the choice of the innermost location of the clouds. If the properties of winds and the parameters ($\epsilon_{\rm B}$, $p$ and $\eta$) are fixed, from Equation (10), we can see that $\nu_{\rm a} \propto \nu_{\rm m} \propto B$. From Equation (5), $B \propto \sqrt{\rho_{\rm wind}}$. The velocity of wind is assumed to be non-variable as introduced above, therefore, the mass flux conservation of wind results in $\rho_{\rm wind} \propto r^{-2}$. Thus, we have $\nu_{\rm a} \propto r^{-1}$. For non-variable wind velocity, the location of wind head (where radio emission is produced) $r \propto t$. Therefore, we have $\nu_{\rm a} \propto t^{-1}$ as shown by the solid line in the left panel of Figure \ref{fig:SED7}. The time evolution of the peak frequency of AT2019dsg (colored dots) follows very well the model predicted pattern.

\begin{figure}
\begin{center}
\includegraphics[width=0.45\textwidth]{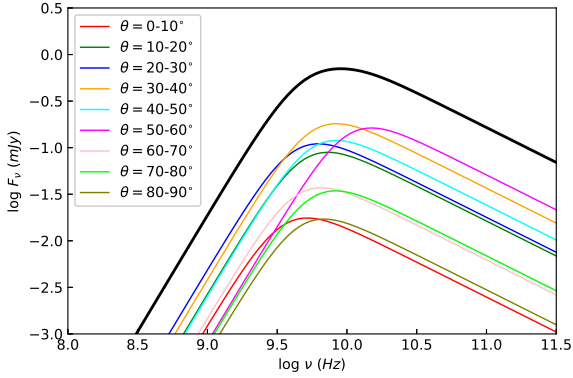}
\caption{Analysis about the contribution of different angular bins of wind (see Section 2.1) to the spectrum for AT2019dsg. The black line shows the SED at 161 day after the outburst date. The colored lines show the contribution from different bins. }
\label{fig:SED7_analysis}
\end{center}
\end{figure}

\begin{figure}
\begin{center}
\includegraphics[width=0.45\textwidth]{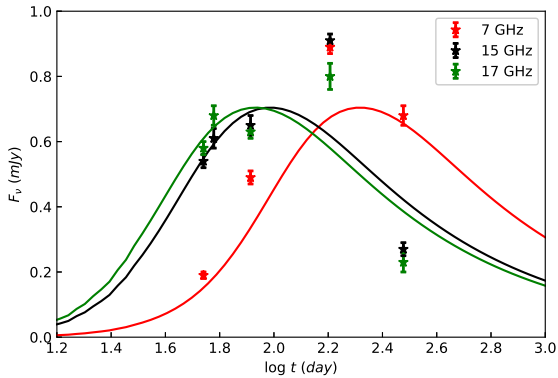}
\caption{Comparison of wind-cloud model predicted light curve (solid lines) and observations (stars) for AT2019dsg at 7GHz (red), 15GHz (black) and 17GHz (green). The observational data of AT2019dsg are from Cendes et al. 2021.}
\label{fig:lightcurve7}
\end{center}
\end{figure}

From Equation (8), for fixed parameters ($k_{\rm bow}$, $c_f$, $\epsilon_{\rm e}$, $\epsilon_{\rm B}$, and $\eta$) and wind properties, $L_{\nu_{\rm m}}$ does not vary with time. As introduced above, $\nu_{\rm a} \propto \nu_{\rm m}$. Therefore, from Equation (12), the peak radio luminosity $L_{\nu_{\rm a}}$ is predicted to be non-evolving with time by our model. As shown by the colored solid lines in the right panel of Figure \ref{fig:SED7}, the model predicted peak flux density ($F_{\nu, \rm a}$) does not vary with time. The peak frequency evolves from high to low frequencies. The time-variation of the peak flux density of AT2019dsg (colored dots) is very small, which is consistent with that predicted by the wind-cloud interaction model.

We analyze the contribution of wind in each angular bin (see Section 2.1) to the SED by plotting Figure \ref{fig:SED7_analysis}. We take the SED at 161 day after the outburst as an example. It is clear that the most important contribution is from wind in angular bin 4 ($30^\circ - 40^\circ$). This is because the kinetic power of wind in bin 4 is largest among all the bins (see the middle-right panel of Figure \ref{fig:wind}). The least contributions are from bins 1 and 9. This is because the kinetic powers of winds in these two bins are the smallest (see the middle-right panel of Figure \ref{fig:wind}). Therefore, the contribution of wind in each angular bins to the SED depends on the kinetic power of wind. The contribution increases with the increase of wind power.

Winds in different angular bins have different velocity (see Figure \ref{fig:wind}). The fastest wind interacts with clouds first. Therefore, the radio emission is first generated by the interaction from the fastest wind in bin 1. However, the kinetic power of winds in bin 1 is not the largest as introduced above. Therefore, the most contributing region shifts from bin 1 to the bin which has the largest wind power. The shift makes the small discontinuities in the curve of $\nu_{\rm a}$ in the left panel of Figure \ref{fig:SED7}.

We show the comparison of model predicted light curve (solid lines) and observations (stars) for AT2019dsg in Figure \ref{fig:lightcurve7}. Light curves are shown for frequencies at 7GHz (red), 15GHz (black) and 17GHz (green). The model predicted emission flux ($F_\nu$) first increases and then decreases with time. The reason for such an evolution pattern is as follows. At early evolution stage, the chosen frequencies (7GHz, 15GHz and 17GHz) are much smaller than $\nu_{\rm a}$. The emission fluxes at these frequencies are much lower than the non-evolution $F_{\nu_{\rm a}}$. With time evolution, $\nu_{\rm a}$ decreases as $t^{-1}$. The chosen frequencies get closer to $\nu_{\rm a}$, and correspondingly the emission fluxes get larger. When the chosen frequencies become equal to $\nu_{\rm a}$, the emission fluxes achieve the maximum value ($F_{\nu_{\rm a}}$). Afterwards, the chosen frequencies become larger than $\nu_{\rm a}$. With time evolution, the radio emission fluxes at the chosen frequencies decrease. The observed emission fluxes roughly follow the model predicted light curve. In the wind-CNM model, the light curve depends on the radial structure of the CNM. If the number density of CNM decreases with radius as $n \propto r^{-2}$, both $\nu_{\rm a}$ and $F_{\nu_{\rm a}}$ have the same time evolution pattern as those predicted in the wind-cloud interaction model (see the Appendix A in \cite{Matsumoto2021}). If the CNM has a distribution of $n \propto r^{-1}$, both $\nu_{\rm a}$ and $F_{\nu_{\rm a}}$ (which depend on $p$) have quite different time evolution pattern from those predicted in wind-cloud interaction model (see the Appendix A in \cite{Matsumoto2021}). The evolution of the light curve of AT2019dsg seems require a $n\propto r^{-2}$ distribution of the CNM in the wind-CNM model.

\begin{figure*}
\begin{center}
\includegraphics[width=0.45\textwidth]{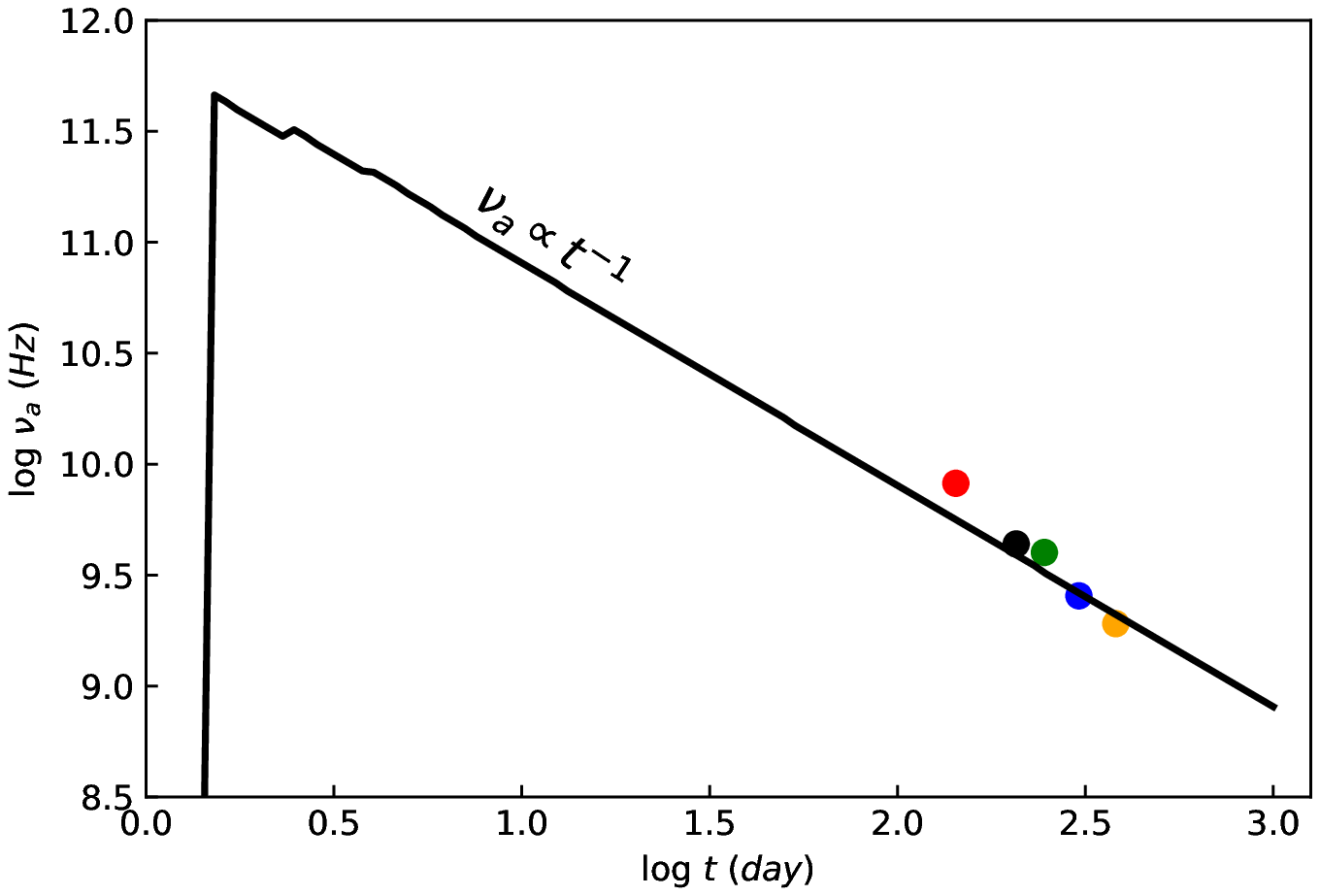}
\includegraphics[width=0.45\textwidth]{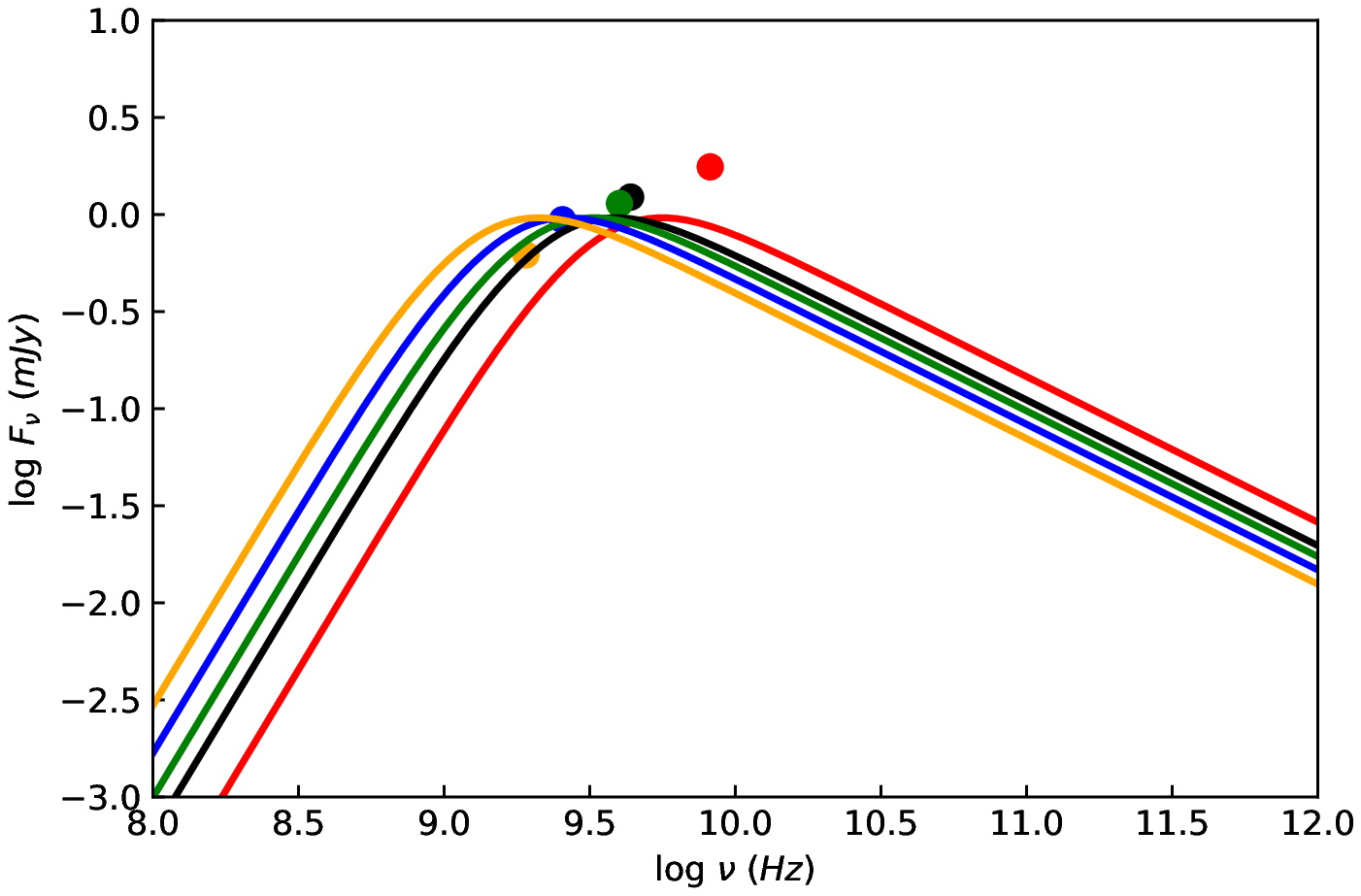}\\
\caption{Time evolution of the peak frequencies (left panel) and SED (right panel). In both panels, the colored dots are for ASASSN-14li. The dots colored red, black, green, blue, and yellow correspond to 143 day, 207 day, 246 day, 304 day and 381 day after the optical discovery date. In the left panel, the solid line is the model predicted evolution of the peak frequency. Note that the peak frequency shown in the left panel is equivalent to the SSA frequency. In the right panel, the colored solid lines are the model predicted SEDs at different snapshot after outburst. The colored dots in the right panel show the peak flux density $F_{\nu, \rm a}$. The details for the parameters of the model are introduced in Section 3.2.}
\label{fig:SED6}
\end{center}
\end{figure*}

\subsection{ASASSN-14li}
ASASSN-14li located in the nucleus of the nearby galaxy PGC 043234 was discovered as a transient on 2014 November 22. ASASSN-14li was confirmed to be a TDE by extensive optical, UV, and X-ray follow-up observations (\cite{Miller2015}; \cite{Holoien2016a}). The redshift of this source $z=0.0206$. The central massive black hole has mass $M_{\rm BH} \sim 10^6M_\odot$.

The transient radio emission from ASASSN-14li was reported independently by \cite{Van2016} and \cite{Alexander2016}. By fitting the multi-frequency radio observations of successive epochs, the peak radio frequency ($\nu_{\rm a}$), the peak flux density ($F_{\nu, \rm a}$), and their time-evolution are obtained (\cite{Alexander2016}).

For ASASSN-14li, the black hole mass $M_{\rm BH} \sim 10^6 M_\odot$. When we calculate the radio emission based on the TDE wind-cloud interaction model, we use the wind properties of model M6 (see the top panels of Figure \ref{fig:wind}). As for the case of AT2019dsg, we adopt parameters $c_f = 0.2$, $\epsilon_{\rm e} = \epsilon_{\rm B} =0.1$, $p=2.5$, and $\eta = 10^{-3}$. We also assume that the clouds are distributed outside $10^{0}$ light days away from the black hole.

Figure \ref{fig:SED6} shows the time evolution of the peak frequency $\nu_{\rm a}$ (left panel), the peak flux density $F_{\nu, \rm a}$ and SED (right panel). As in Figure \ref{fig:SED7}, the solid line in the left panel is the model predicted evolution of the peak frequency. The fastest winds have a velocity of $\sim 0.7 c$ (see the top left panel of Figure \ref{fig:wind}). We assume that the clouds are distributed outside $10^{0}$ light-days. It takes $\sim 1.4$ days for the TDE winds to arrive at a distance of $10^{0}$ light-days to interact with the clouds, which produces radio emission. The radio emission arises after 1.4 days after the outburst as shown by the solid line in the left panel of Figure \ref{fig:SED6}. As introduced above, for the fixed model parameters of ($\epsilon_{\rm B}$, $p$ and $\eta$), the model predicted $\nu_{\rm a} \propto t^{-1}$ as shown by the solid line in the left panel of Figure \ref{fig:SED6}. The time evolution of the peak frequency of ASASSN-14li (colored dots) follows very well the model predicted pattern.
The time-variation of the peak flux density of ASASSN-14li (colored dots) is also consistent with that predicted by the model.

\begin{figure}
\begin{center}
\includegraphics[width=0.45\textwidth]{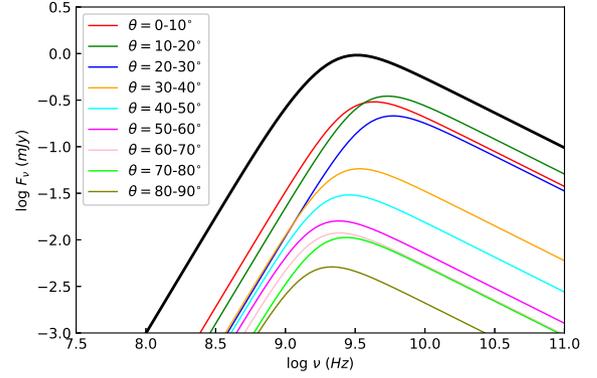}
\caption{Analysis about the contribution of different angular bins of wind (see Section 2.1) to the spectrum for ASASSN-14li. The black line shows the SED at 246 day after the outburst date. The colored lines show the contribution from different bins. }
\label{fig:SED6_analysis}
\end{center}
\end{figure}

We analyze the contribution of wind in each angular bin (see Section 2.1) to the SED by plotting Figure \ref{fig:SED6_analysis}. We take the SED at 246 day after the outburst as an example. It is clear that the most important contribution is from wind in angular bin 2 ($10^\circ - 20^\circ$). This is because the kinetic power of wind in bin 2 is largest among all the bins (see the middle-left panel of  Figure \ref{fig:wind}). The least contribution is from bins 9. This is because the kinetic power of winds in bin 9 is the smallest (see the middle-left panel of Figure \ref{fig:wind}). As introduced above, the small discontinuities in the curve of $\nu_{\rm a}$ in the left panel of Figure \ref{fig:SED6} is due to the shift of the most contributing angular bin.

\begin{figure}
\begin{center}
\includegraphics[width=0.45\textwidth]{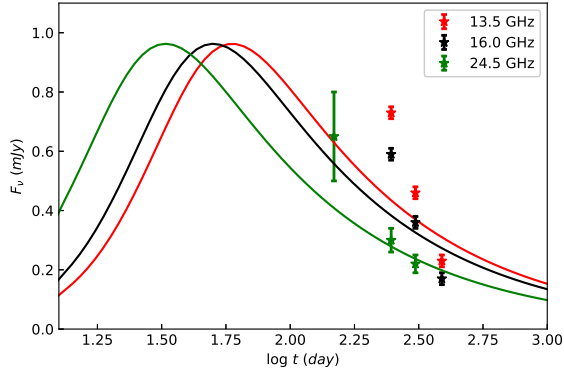}
\caption{Comparison of wind-cloud model predicted light curve (solid lines) and observations (stars) for ASASSN-14li at 13.5GHz (red), 16GHz (black) and 24.5GHz (green). The observational data of ASASSN-14li are from Alexander et al. 2016.}
\label{fig:lightcurve6}
\end{center}
\end{figure}

We show the comparison of model predicted light curve (solid lines) and observations (stars) for ASASSN-14li in Figure \ref{fig:lightcurve6}. Light curves are shown for frequencies at 13.5GHz (red), 16GHz (black) and 24.5GHz (green). It can be seen that the evolution trends are roughly consistent with each other. We note that the model predicted declining rate is not same as observations. More physics may be needed in the wind-cloud interaction model to explain the observations better. In the wind-CNM model for ASASSN-14li, the required density profile for CNM is steeper than $n\propto r^{-2}$ (e.g., \cite{Alexander2016}).

\subsection{ASASSN-15oi and CSS161010}
ASASSN-15oi was discovered in optical bands on 2015 August 14 (\cite{Holoien2016b}) and classified as a TDE. This source has a redshif $z=0.0484$ with a supermassive black hole with mass $M_{\rm BH} \sim 10^6 M_\odot$ (\cite{Gezari2017}; \cite{Holoien2018}). There are only four successive observation epochs for ASASSN-15oi which show the peak radio freqency. Also, the time interval between two adjacent observation epochs (with peak frequency) is very short (\cite{Horesh2021}). Therefore, it is hard to give a time-evolution pattern of the peak frequency. We do not do detailed comparisons between the TDEs winds-cloud interaction model to the radio observations of ASASSN-15oi. However, we note that the luminosity $\nu L_{\nu}$ at 5 GHz after 200 days the optical discovery is $2-3 \times 10^{38}$ erg/s (\cite{Horesh2021}). By using the TDE winds properties of model M6 and setting $\epsilon_{\rm e} = \epsilon_{\rm B} =0.1$, $p=2.5$, and $\eta = 10^{-3}$, we find that the TDE wind-cloud interaction model predicted luminosity at 5 GHz after 200 days of ourburst is $\nu L_\nu \sim 5 \times 10^{37}$ erg/s. The model predicted luminosity at 5GHz is roughly consistent with observations.

The host galaxy of CSS161010 is a dwarf galaxy with a total stellar mass $\sim 10^7M_\odot$ (\cite{Coppejans2020}). The black hole is estimated to be $M_{\rm BH} \sim 10^3 M_\odot$. We can not use the wind-cloud interaction model to explain the observations of CSS161010. The reasons are twofold. First, the properties of the winds from the disruption of a star by an intermediate mass black hole are unknown. Second, for a dwarf galaxy, the presence/absence and the properties of clouds are not known. However, we note an interesting point that for CSS161010 the peak frequency roughly evolves as $\nu_{\rm a} \propto t^{-1}$ (\cite{Coppejans2020}).

\subsection{Effects of varying parameters}
We have parameters $k_{\rm bow}$, $c_f$, $\epsilon_{\rm B}$, $\epsilon_{\rm e}$, $R_{\rm c}/r$, and the power law index of NTe $p$. The value of $v_{\rm wind}/k_{\rm bow}$ represents the adiabatic expansion velocity of the shocked wind. The value $k_{\rm bow} = 10$ is from the detailed simulations studying the colliding of the wind and cloud (\cite{Mou2021}). Thus, there is very little freedom to adjust the value of $k_{\rm bow}$. The power law index of NTe $p$ ($p = 2\alpha+1$ with $\alpha$ being the observed spectral index) can be constrained from observations. Therefore, we also do not study the effect of varying $p$.

\begin{figure*}
\begin{center}
\includegraphics[width=0.45\textwidth]{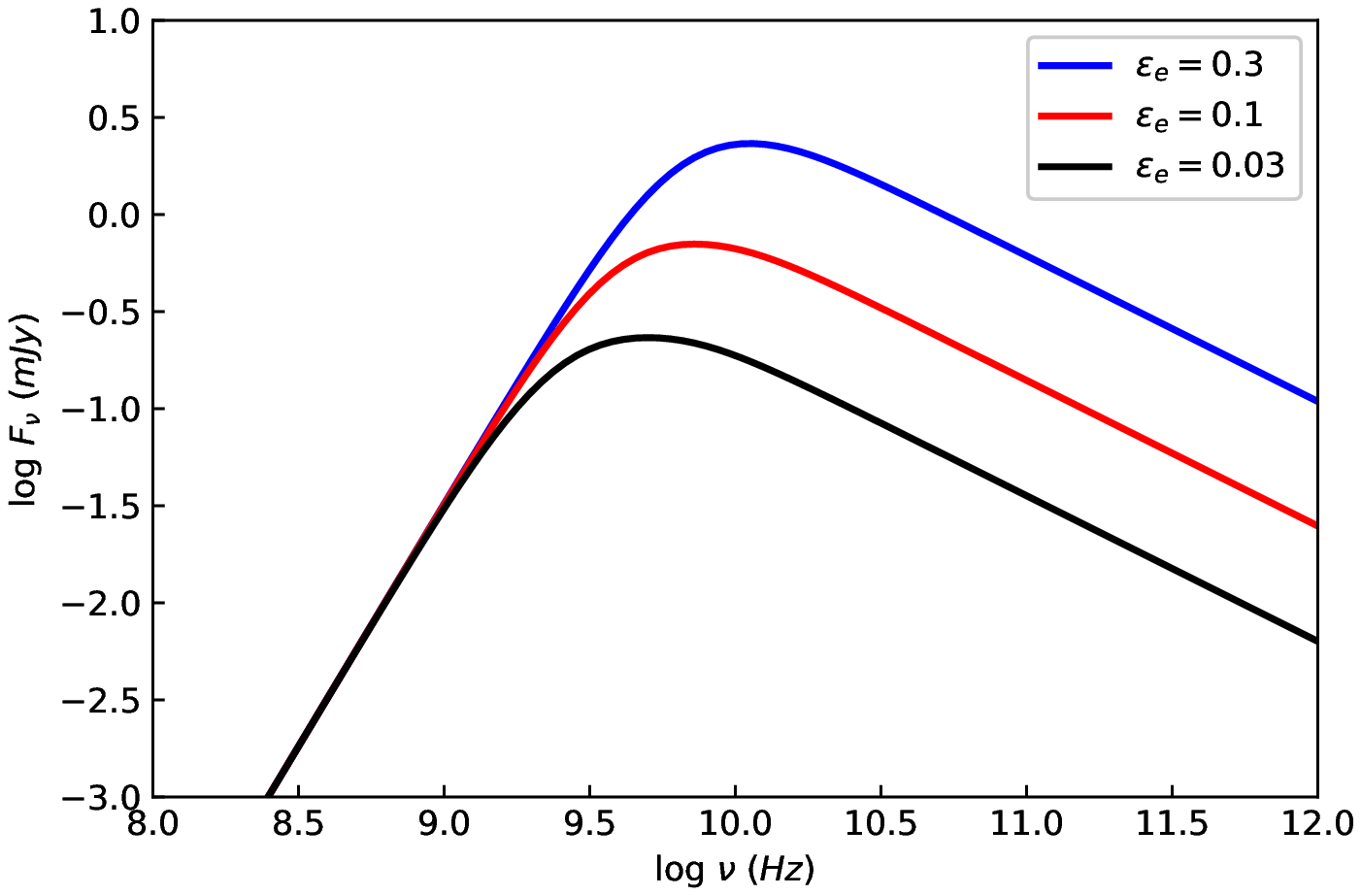}
\includegraphics[width=0.45\textwidth]{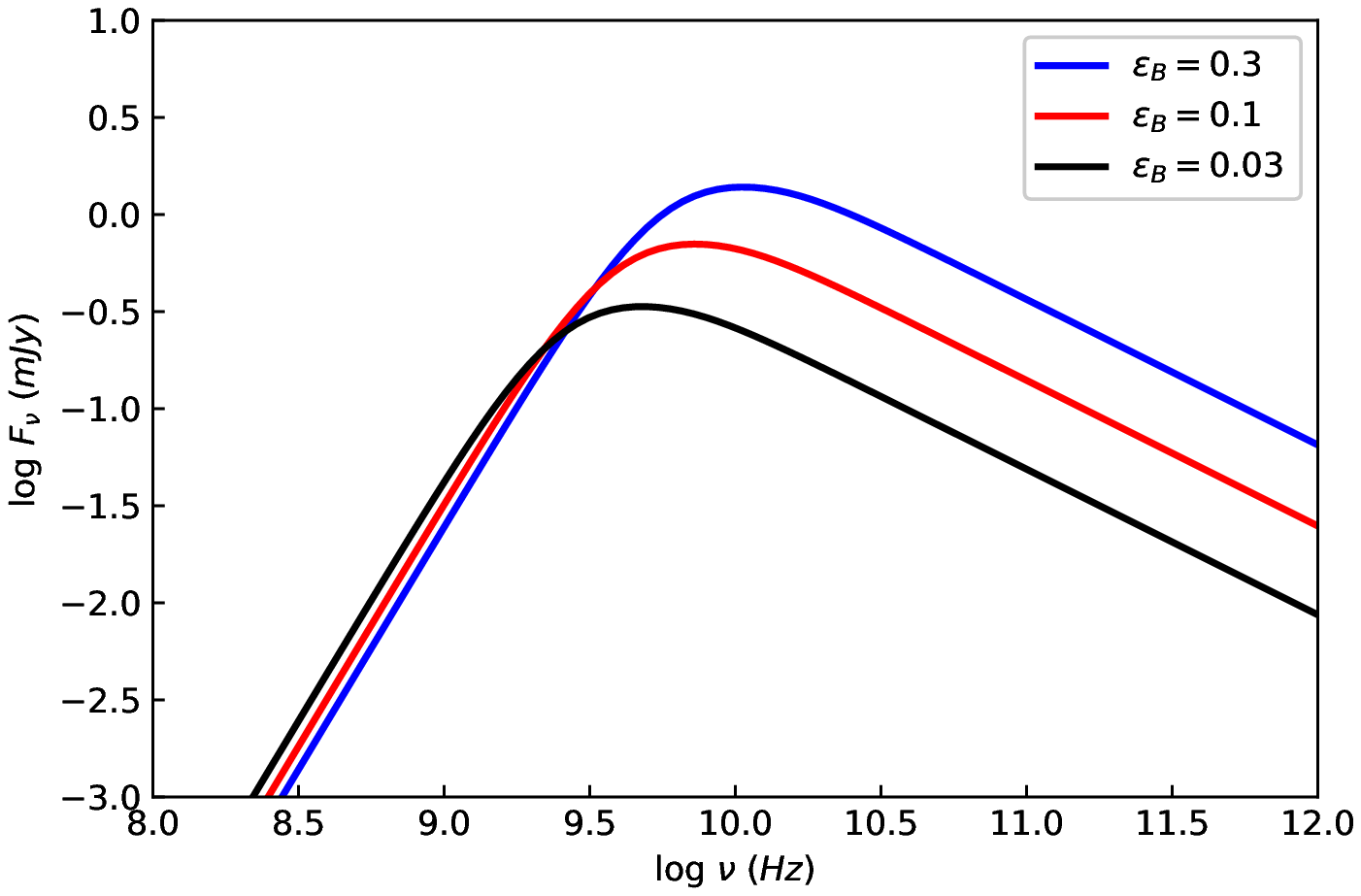}\\
\includegraphics[width=0.45\textwidth]{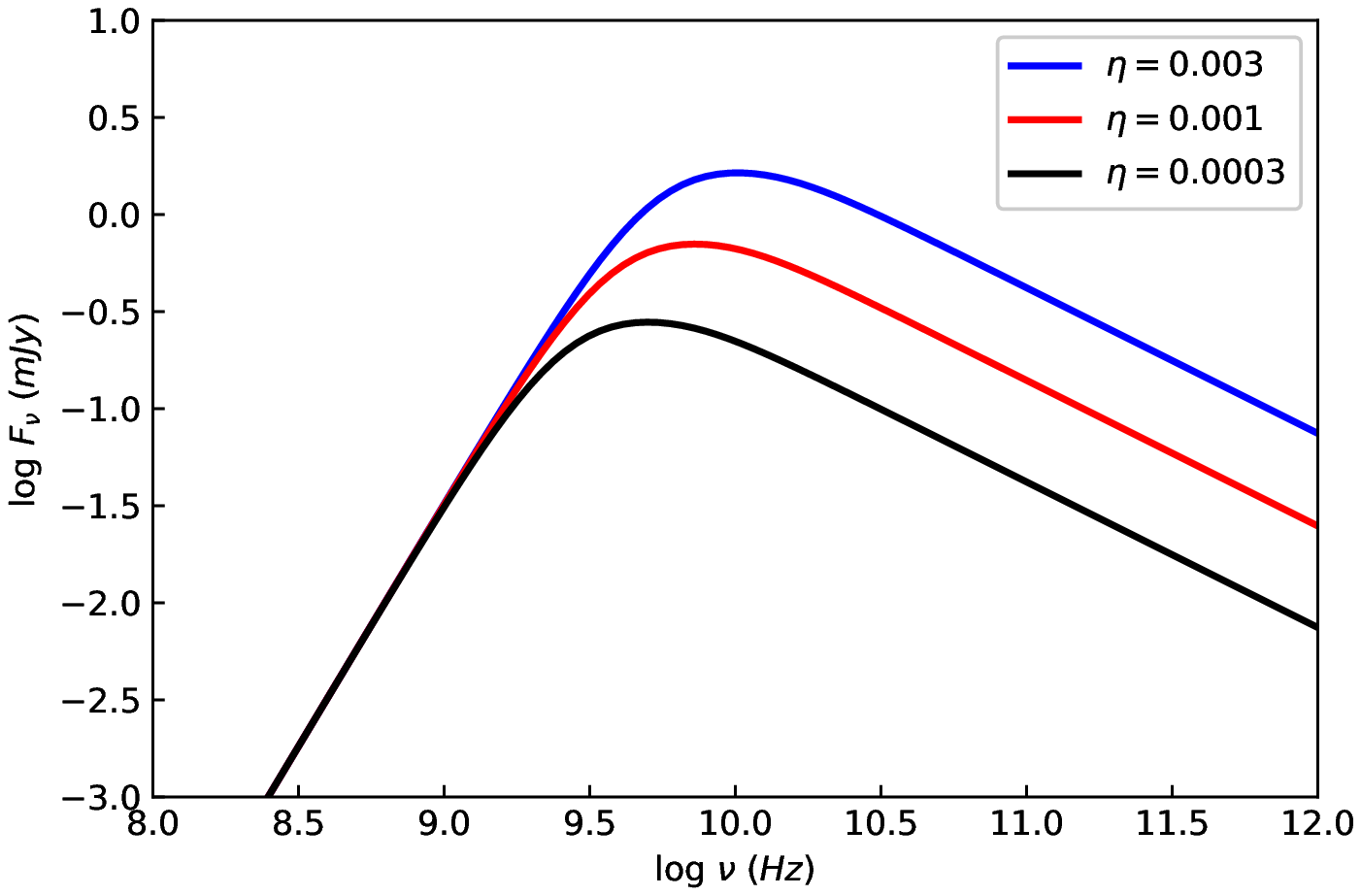}
\includegraphics[width=0.45\textwidth]{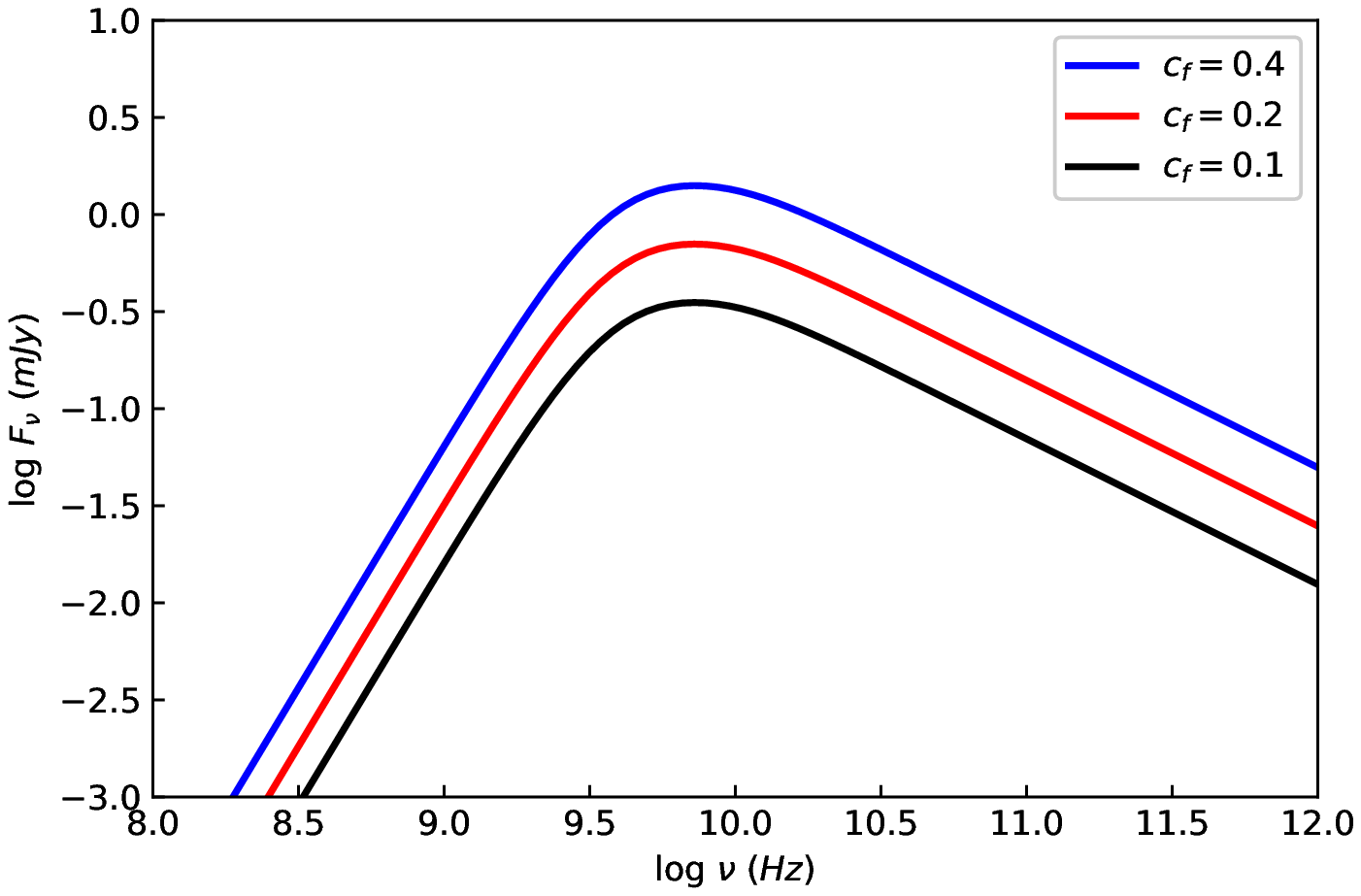}
\caption{Effects of varying parameters. We set the black hole mass $M_{\rm BH} = 10^7 M_\odot$. As in section 3.1 of studying the emission of AT2019dsg, we set that the clouds are distributed outside $10^{1/2}$ light days away from the black hole. The figure shows the results of the SED at 200 days after the TDEs outburst. In all the panels, we set $k_{\rm bow} = 10$ and $p = 2.5$. The upper-left panel shows the effects of varying $\epsilon_{\rm e}$ with $\epsilon_{\rm B} = 0.1$, $c_f=0.2$ and $R_{\rm c}/r = 10^{-3}$. The upper-right panel shows the effects of varying $\epsilon_{\rm B}$ with $\epsilon_{\rm e} = 0.1$, $c_f=0.2$ and $R_{\rm c}/r = 10^{-3}$. The bottom-left panel shows the effects of varying $\eta = R_{\rm c}/r$ with $\epsilon_{\rm e} = \epsilon_{\rm B} = 0.1$ and $c_f = 0.2$. The bottom-right panel shows the effects of varying $c_f$ with $\epsilon_{\rm e} = \epsilon_{\rm B} = 0.1$ and $R_{\rm c}/r = 10^{-3}$. }
\label{fig:parameter}
\end{center}
\end{figure*}

We study the effects of varing $c_f$, $\epsilon_{\rm B}$, $\epsilon_{\rm e}$ and $R_{\rm c}/r$. When we do this study we set the black hole mass $M_{\rm BH} = 10^7 M_\odot$. As in section 3.1 of studying the emission of AT2019dsg, we set that the clouds are distributed outside $10^{1/2}$ light days away from the black hole. Figure \ref{fig:parameter} shows the results of the SED at 200 days after the TDEs outburst.

From the top-left panel of Figure \ref{fig:parameter}, we see that the value of $\nu_{\rm a}$ increases with $\epsilon_{\rm e}$. This is because that with other parameters fixed, the increase of $\epsilon_{\rm e}$ results in the increase of the number of non-thermal electrons. With the emission area fixed, the increase of the number of NTe can result in $\nu_{\rm a}$ getting larger. Of course, because the value of $\epsilon_{\rm e}$ in the mildly relativistic regime ($v_{\rm wind} > 0.3 c$) also affects the minimum value of the electron Lorentz factor $\gamma_{\rm m}$, the value of $\nu_{\rm a}$ do not correlate linearly with $\epsilon_{\rm e}$. Anyway, the value of $\nu_{\rm a}$ correlates positively with $\epsilon_{\rm e}$ (see also Equation 8 of MOU22). The luminosity in the optically thick region can not be affected by $\epsilon_{\rm e}$. The reason is as follows. The emitting area is $\sim R_{\rm c}^2$, which does not vary. Optically thick synchrotron emission flux per unit area is not affected by the number density of NTe. Therefore, the emission flux for the optically thick regime ($\nu < \nu_{\rm a}$) do not vary with $\epsilon_{\rm e}$. However, for the optically thin regime, the increase of the number density of NTe can increase the radiation flux.

The absorption coefficient of synchrotron emission $\alpha_\nu \propto B^{-1} \nu_{\rm m}^{(p+4)/2} \nu^{-(p+4)/2}$ and $\nu_{\rm m} \propto B$. Therefore, $\alpha_\nu \propto B^{(p+2)/2} \nu^{-(p+4)/2} $. The requirement that $\tau_{\nu_{\rm a}} = \alpha_\nu l = 1$ (with $l$ being the unchanged size of cloud), the increase of the magnetic field can result in the increase of $\nu_{\rm a}$, which can be seen from the top-right panel of Figure \ref{fig:parameter}. For optically thin synchrotron emission, from Equation (7), the radiation luminosity $\nu L_\nu$ is proportional to $B^2$. Therefore, in the optically thin regime, the radiation flux positively correlates with $\epsilon_{\rm B}$ as shown in the upper-right panel of Figure \ref{fig:parameter}. In the optically thick regime, the emission flux per unit area is weakly dependent on the magnetic field ($\propto B^{-1/2}$). Therefore, the radiation flux in the optically thick regime is just slightly affected.

We show the effects of varying $\eta = R_{\rm c}/r$ in the bottom-left panel of Figure \ref{fig:parameter}. With the covering factor of clouds fixed, the increase (or decrease) of $\eta$ means that the size of each cloud increases (or decreases) and the number of clouds decreases (or increases). The value of $\nu_{\rm a}$ is determined by $\alpha_{\nu} l = \alpha_{\nu} R_{\rm c} =1$. Because $\alpha_\nu \propto \nu^{-(p+4)/2}$, with the increase (or decrease) of $R_{\rm c}$, the value of $\nu_{\rm a}$ increases (or decreases) as shown in the figure. The increase (or decrease) of $R_{\rm c}$ results in increase (or decrease) of adiabatic expansion timescale $t_{\rm ad}$. From Equation (4), we see that the increase (or decrease) of $t_{\rm ad}$ increases (or decreases) the number of NTe. Therefore, for the optically thin regime ($\nu > \nu_{\rm a}$), the radiation flux increases (or decreases) with the increase (or decrease) of $\eta$. However, in the optically thick regime ($\nu < \nu_{\rm a}$), the radiation flux is not affected. The reason is as follows. The covering factor is not changed, therefore, the total radiation area is not changed. For the optically thick radiation, the radiation flux per unit area is only affected by magnetic field, which is not changed. Therefore, the radiation flux in the optically thick regime is not affected.

With the size of the clouds fixed, the change of covering factor means that the number of clouds changes. For each cloud, the radiation properties is not changed. Therefore, the increase or decrease of number of cloud will make the SED move upwards or downwards as shown in the bottom-right panel of Figure \ref{fig:parameter}.

From the analysis above, we can see that the parameters $\epsilon_{\rm e}$, $\epsilon_{\rm B}$ and $\eta$ have degeneracies. We can obtain very similar results by increasing (or decreasing) one or two of the three parameters and simultaneously decreasing (or increasing) the last parameter.

\section{Summary and Discussions}
The winds generated in TDEs can interact with the condensed cloud surrounding the central black hole, produce bow shocks, accelerate electrons, and produce radio emission (MOU22). In MOU22, due to the poor knowledge of the TDE winds, the authors can only predict the time-evolution pattern of the peak radio emission frequency. However, the values of both the peak radiation frequency and the peak luminosity can not be calculated.

In this paper, we study the TDEs wind-cloud interaction model. The properties of winds are given by the radiation hydrodynamic simulations of super-Eddington circularized accretion flow in TDEs (BU22). We can calculate the peak radio frequency, the luminosity at the peak frequency, and their time-evolutions based on the TDEs wind-cloud interaction model. We compare the model predicted radio emission properties to two radio TDEs, namely, AT2019dsg and ASASSN-14li. We find that the model predicted  peak radiation frequency, the luminosity at peak frequency, and their time evolution can be well consistent with those in TDEs AT2019dsg and ASASSN-14li. This indicates that in these two radio TDEs, the wind-cloud interaction mechanism may be responsible for the radio emission.

The TDE wind-CNM model predicts that for AT2019dsg, the velocity of wind is 0.12$c$ (\cite{Stein2021}) or 0.07$c$ (\cite{Cendes2021}). The black hole mass of AT2019dsg $M_{\rm BH} \sim 10^7 M_\odot$. The simulations (BU22) show that the winds from a circularized super-Eddington accretion flow in TDEs with black hole mass $10^7M_\odot$ is in the range 0.1-0.4$c$ (see the bottom left panel of Figure \ref{fig:wind}). For ASASSN-14li with black hole mass $M_{\rm BH} \sim 10^6 M_\odot$, the wind-CNM model suggests a wind velocity of 0.04-0.12$c$ (\cite{Alexander2016}). The simulations of BU22 find that the winds generated in circularized accretion flow in TDEs with $M_{\rm BH} = 10^6 M_\odot$ is in the range 0.1-0.7$c$ (see the top left panel of Figure \ref{fig:wind}). The velocity of TDE winds suggested by the wind-CNM model is much lower than that found by our simulations. For ASASSN-14li, an ionized wind with a velocity of $0.2c$ is detected (\cite{Kara2018}), which is much consistent with our simulation result in BU22.

In this paper, we assume that the wind-cloud interaction process has very little effects on dissipating wind kinetic energy and decelerating wind. Thus, we assume that the velocity of wind does not change after interaction with clouds. We do some analysis to illustrate the rationality of this assumption. The energy needed to power the radio emission is given by some previous papers. For example, \cite{Cendes2021} find that to power the radio emission of AT2019dsg, the energy of electrons is in the range of $10^{47}-4\times 10^{48}$ erg. Numerical simulations (BU22; \cite{Curd2019}; \cite{Dai2018}) found that the kinetic power of wind in the peak outburst stage is roughly $2-4\times 10^{44}$ erg/s. If we assume that the peak stage of a TDE outburst can last for 20 days, the total kinetic energy of wind is 20 days $\times$ $8.64\times10^4$ second/day $\times 4 \times 10^{44}$ erg/s, which is roughly $7\times10^{50}$ erg. The kinetic energy of wind (obtained by simulations) can be more than 2 orders of magnitude higher than the energy of electrons (\cite{Cendes2021}). Thus, the wind energy dissipation would be not important. The deceleration of wind by clouds would be not important.

For some TDEs, the radio luminosity presents fluctuations (\cite{Perlman2022}) with two peaks during years of monitoring. Regarding ASASSN-15oi, this event shows 2 radio peaks (\cite{Horesh2021}). The first peak is detected $\sim 180$ days after optical discovery. A very bright delayed second radio peak is detected $\sim 1400$ days after optical discovery. If we use the wind-cloud interaction model to explain the two peaks phenomenon, we need two separated cloud regions. If we take a value of wind $v_{\rm wind} = 0.5c$, we can calculate the locations of the two separated clouds regions. The inner clouds region is around $v_{\rm wind}t \sim 0.07 {\rm parsec}$ (with $t$ being 180 days). The outer clouds region is around $v_{\rm wind}t \sim 0.6 {\rm parsec}$ (with $t$ being 1400 days). The physical reason to form separated dense cloud regions is not known. There are some observations, which show separated dense gas regions in our Galaxy center. Observations of Galaxy center show a very dense circumnuclear disk (CND) at the parsec scale (\cite{Zhao2016}) and an inner `mini-spiral' (\cite{Tsuboi2016}). The two peaks phenomenon can also be explained by the wind-CNM model. A non-monotonic radial CNM structure can be produced by stellar activities (e.g., \cite{Generozov2017}). The multi-peak radio emission can be produced when the wind interacts with CNM, which has a non-monotonic radial structure.  A delayed radio emission is also detected in AT2018hyz (\cite{Cendes2022}), which could be explained by a delayed wind (\cite{Cendes2022}) or an off-axis jet (\cite{MatsumotoPiran2022}) in the wind-CNM model.
    
In the `circularization' process of the fallback debris, winds can be generated by the stream-stream collision (\cite{Jiang2016}; \cite{Lu2020}). In future, it is interesting to restudy the wind-cloud interaction model by adopting the properties of winds generated in the stream-stream collision process.

\section*{Acknowledgments}
D. Bu is supported by the Natural Science Foundation of China (grants 12173065, 12133008, 12192220, 12192223) and the science research grants from the China Manned Space Project (No. CMS-CSST-2021-B02). L. Chen is supported by the Natural Science Foundation of China (grant 12173066) and Shanghai Pilot Program for Basic Research – Chinese Academy of Science, Shanghai Branch, JCYJ-SHFY-2021-013. G. Mou is supported by the Natural Science Foundation of China (grant 1183007). E. Qiao is supported by the National Natural Science Foundation of China (grant 12173048) and NAOC Nebula Talents Program. X. Yang is supported by the Natural Science Foundation of China (grant 11973018).
\section*{Data availability}
The data underlying this article will be shared on reasonable request to the corresponding author.

\appendix
\section{The definition of TDE wind head}
In BU22, we simulate the circularized accretion process of stellar debris in TDEs. The TDE winds can be produced in the super-Eddington accretion phase. In BU22, the TDE winds move outward in an almost `vacuum' circumstance, because the ambient gas density is set to be several orders of magnitude lower than the density of the TDE winds. The ambient gas can not affect the properties of TDE winds at all. In the simulations of BU22, there is no study about the interaction between TDE winds and ambient circumstance (circumnuclear medium or dense clouds).

\begin{figure}
\begin{center}
\includegraphics[scale=0.42]{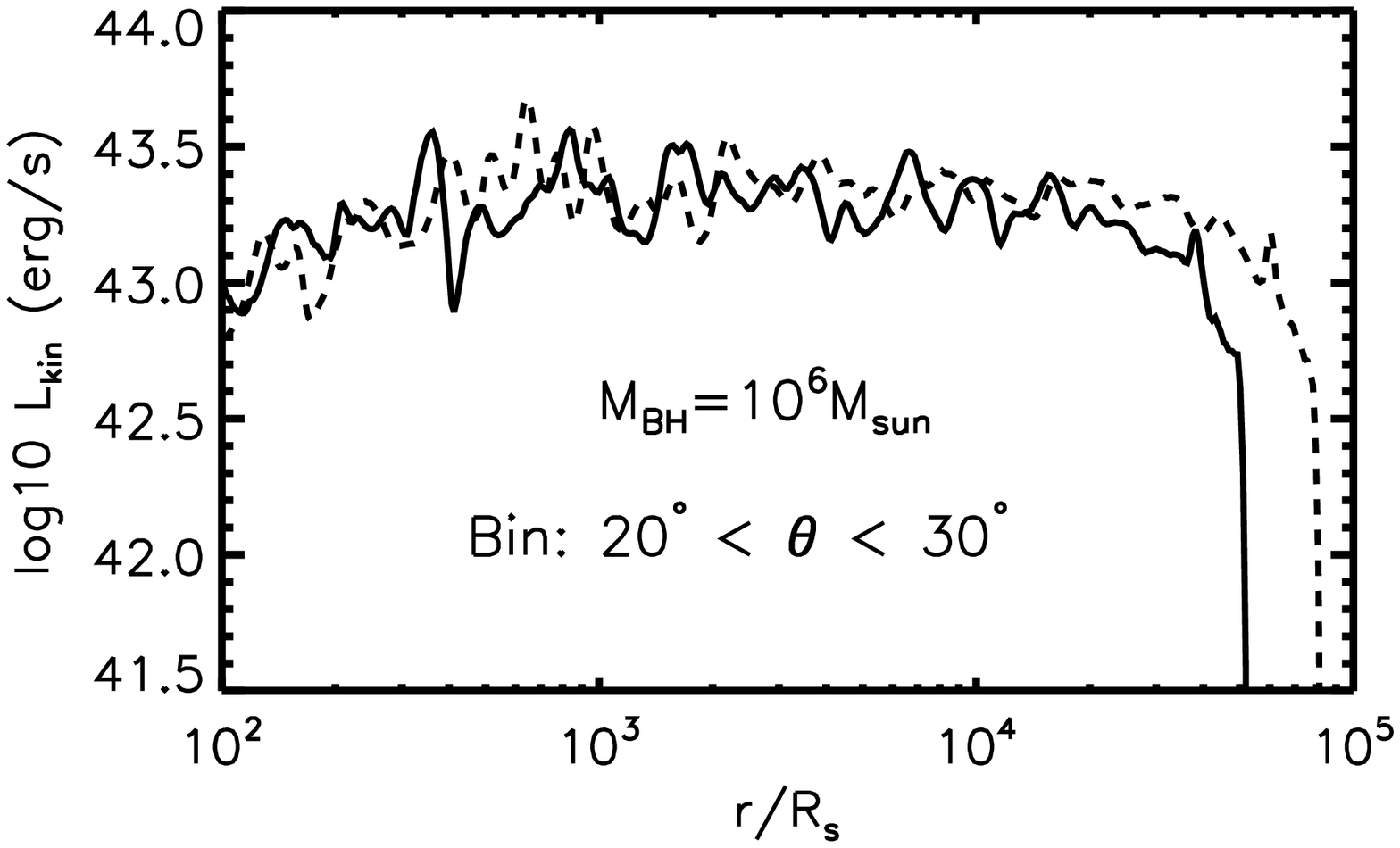}\hspace*{0.5cm}
\hspace*{0.5cm} \caption{Radial profile of the TDE winds kinetic power for the bin of $20^\circ < \theta < 30^\circ$ for model M6 with $M_{\rm BH}=10^6 M_\odot$. The solid and dashed lines are corresponding to time $=21$ day and $32$ day, respectively. In the label of the horizontal axis, $R_s$ is the Schwarzschild radius. \label{Fig1:A1}}
\end{center}
\end{figure}

In BU22, the TDE winds move in an almost `vacuum'. If we plot the radial distribution of wind properties at a snapshot, we will find a cutoff at a certain radius. The cutoff radius is the location the wind has arrived at this snapshot. Outside the cutoff radius, it is `vacuum'. We take the simulation M6 in BU22 as an example. Figure \ref{Fig1:A1} plots the radial profiles of the TDE winds kinetic power for the bin of $20^\circ < \theta < 30^\circ$ for model M6 with $M_{\rm BH}=10^6 M_\odot$. The solid and dashed lines are corresponding to time $=21$ day and $32$ day, respectively. At $21$ day, the cutoff radius is $5 \times 10^4 R_{\rm s}$ (with $R_{\rm s}$ being the Schwarzschild radius). At $32$ day, the cutoff radius moves to $8 \times 10^4 R_{\rm s}$.  Because the winds keep moving outwards, the cutoff radius will increase with time. The method we define the wind head is as follows. We look for the first maximal value on the radial profile from outside to inside. The radial location where the first maximal value is located is defined as the wind head. For example, in Figure \ref{Fig1:A1}, at time = $21$ day, the wind head is located at $3.5 \times 10^4 R_{\rm s}$; at time = $32$ day, the wind head is located at $6 \times 10^4 R_{\rm s}$. From Figure \ref{Fig1:A1}, although the wind head moves outwards, the kinetic power at the wind head almost does not vary. The reason is that the wind has a velocity significantly larger than the escape velocity, therefore, the gravity can hardly affect the movement of the wind. After finding the location of wind head, we can calculate the properties of wind head shown in Figure \ref{fig:wind}.

\bibliographystyle{rasti}
\bibliography{ref} 








\bsp	
\label{lastpage}
\end{document}